\documentclass[usenatbib]{mn2e}
\usepackage{graphicx, amssymb, aas_macros, natbib, bm}

\usepackage{mathptmx}
\usepackage{epsfig}
\usepackage{wrapfig}
\usepackage{amsmath}
\usepackage{natbib}
\usepackage{graphicx}
\usepackage{subfigure}
\usepackage{enumitem}

\def\msun{\hbox{${\rm M}_{\odot}$}}
\newcommand{\aco}{\ensuremath{\alpha_{\rm CO}}}
\newcommand{\acom}{\alpha_{\rm CO}}

\newcommand{\ha}{$\mathrm{H\alpha}$}
\newcommand{\re}{$r_{\rm e}$}

\newcommand{\dhst}{3D-HST}

\newcommand{\um}{$\mathrm{\mu m}$}
\newcommand{\sigtot}{\ensuremath{\Sigma_{\rm tot}}}

\newcommand{\htwo}{$\mathrm{H_2}$}
\newcommand{\lco}{$L'_{\rm CO}$}
\newcommand{\tdep}{\ensuremath{t_{\rm dep}}}
\newcommand{\acounit}{\ensuremath{\mathrm{M_\odot~(K~km~s^{-1}~pc^2)^{-1}}}}
\newcommand{\mscinot}[1]{\times 10^{#1}}
\newcommand{\galex}{\textit{GALEX}}

\defcitealias{bolatto2013}{B13}
\defcitealias{sandstrom2013}{S13}

 \title[PHIBSS: \aco{} at $z < 1.5$]{PHIBSS: Exploring the Dependence of the CO-H$\mathbf{{_2}}$ Conversion
  Factor on Total Mass Surface Density at $\mathbf{\emph{z} < 1.5}$}
\author[Carleton et al.]
{Timothy Carleton,$^{1}$\thanks{$\!\!$e-mail: tcarleto@uci.edu}
 Michael C. Cooper,$^{1}$\thanks{$\!\!$e-mail: cooper@uci.edu}
 \newauthor Alberto D. Bolatto,$^{2}$
 Frederic Bournaud,$^{3}$
 Fran\c{c}oise Combes,$^{4}$ 
 Jonathan Freundlich,$^{4}$
\newauthor Santiago Garcia-Burillo,$^{5}$
 Reinhard Genzel,$^{6,7,8}$
 Roberto Neri,$^{9}$ 
 Linda J. Tacconi,$^{6}$ 
\newauthor Karin M. Sandstrom,$^{10}$ 
 Benjamin J. Weiner,$^{11}$ 
 Axel Weiss$^{12}$ \\ \\
 $\!\!^1$Center for Cosmology, Department of Physics and Astronomy, 
 4129 Reines Hall, University of California, Irvine, CA 92697,
 USA \\
 $\!\!^2$Department of Astronomy, University of Maryland, College Park,
 MD 20742, USA \\
 $\!\!^{3}$Service d'Astrophysique, DAPNIA, CEA/Saclay, F-91191 Gif-sur-Yvette
 Cedex, France \\
 $\!\!^{4}$ LERMA, Observatoire de Paris, CNRS, UPMC, PSL 
 Univ., and College de France, Paris, France \\
 $\!\!^{5}$Observatorio Astron\'omico Nacional (OAN-IGN)-Observatorio de
 Madrid, Alfonso XII, 3, 28014 Madrid, Spain \\
 $\!\!^{6}$Max-Planck-Institut f\"{u}r Extraterrestrische Physik (MPE),
 Giessenbachstr., D-85748 Garching, Germany \\
 $\!\!^{7}$Department of Physics, Le Conte Hall, University of California,
 Berkeley, CA 94720, USA \\
 $\!\!^{8}$Department of Astronomy, Campbell Hall, University of California,
 Berkeley, CA 94720, USA \\
 $\!\!^{9}$IRAM, 300 Rue de la Piscine, F-38406 St. Martin d'Heres, Grenoble,
 France \\
 $\!\!^{10}$Center for Astrophysics and Space Sciences, Department of Physics,
 University of California, San Diego, 9500 Gilman Drive, La Jolla, CA
 92093, USA  \\
 $\!\!^{11}$Steward Observatory, 933 North Cherry Avenue, University of Arizona,
 Tucson, AZ 85721, USA \\
 $\!\!^{12}$Max Planck Institut f\"{u}r Radioastronomie (MPIfR), Auf
 dem H\"{u}gel 69, 53121 Bonn, Germany 
}

\begin{document}

\pagerange{\pageref{firstpage}--\pageref{lastpage}} 
\pubyear{2016}

\maketitle

\begin{abstract}
  We present an analysis of the relationship between the CO-H$_{2}$
  conversion factor (\aco{}) and total mass surface density
  ($\Sigma_{\rm tot}$) in star-forming galaxies at $z<1.5$.
  Our sample, which is drawn from the IRAM Plateau de Bure HIgh-$z$
  Blue Sequence Survey (PHIBSS) and the CO Legacy Database for GASS (COLD~GASS), includes
  `normal,' massive star-forming galaxies that dominate the evolution
  of the cosmic star formation rate (SFR) at this epoch and probe the
  $\Sigma_{\rm tot}$ regime where the strongest variation in \aco{} is
  observed.
  We constrain \aco{} via existing CO observations, measurements of
  the star formation rate, and an assumed molecular gas depletion time
  (\tdep{}=$M_{\rm gas}$/SFR) --- the latter two of which establish
  the total molecular gas mass independent of the observed CO
  luminosity.
  For a broad range of adopted depletion times, we find that \aco{} is
  independent of total mass surface density, with little deviation
  from the canonical Milky~Way value.
  This runs contrary to a scenario in which \aco{} decreases as
  surface density increases within the extended clouds of molecular
  gas that potentially fuel clumps of star formation in $z\sim1$
  galaxies, similar to those observed in local ULIRGs.
  Instead, our results suggest that molecular gas, both at $z\sim0$
  and $z\sim1$, is primarily in the form of self-gravitating molecular
  clouds.
  While CO observations suggest a factor of $\sim3$ reduction in the
  average molecular gas depletion time between $z \sim 0$ and $z \sim
  1$, we find that, for typical galaxies, the structure of molecular
  gas and the process of star formation at $z \sim 1$ is otherwise remarkably
  similar to that observed in local star-forming systems.

\end{abstract}

\begin{keywords}
	galaxies: formation, evolution, high-redshift, ISM, star formation, ISM: molecules
	\end{keywords}

\clearpage

\section{Introduction}
\label{sec:introduction}
From $z \sim 2$ to today, the volume-averaged cosmic star formation
rate (SFR) has declined by an order of magnitude, both globally
\citep{madau1996, lilly1996} and at fixed stellar mass
\citep{noeske2007, daddi2007, elbaz2007},
driven by a relatively smooth reduction in the star-forming activity
of all galaxies rather than a dramatic evolution in the prevalence of
mergers or starbursts \citep{reddy2008, magnelli2011, madau2014}.
This global reduction in star formation is potentially precipitated by
either a decrease in the supply of cold gas to galaxies over cosmic
time or a lowering in the rate at which gas is converted into stars
(i.e.~an evolution in star-formation efficiency).
To address this question, the Plateau de Bure High-$z$ Blue Sequence
Survey \citep[PHIBSS;][]{tacconi2010, tacconi2013} measured the
molecular gas masses for a sizable sample of typical star-forming
galaxies at $z>1$, finding that cold gas fractions at intermediate
redshift are generally a factor of $3$ larger, with only a modest
evolution in the average star-formation efficiency.
Newer studies making use of dust-based gas measurements 
\citep{genzel2015, santini2014} and
additional CO data from PHIBSS2 \citep[][Tacconi et al.~in prep,]{genzel2015} 
arrive at similar conclusions \cite[see also][]{papovich2016}.
These results suggest that evolution in the cosmic star formation
space density stems chiefly from a decrease in the supply of cold gas,
with no evidence for substantial evolution in the physics of star
formation over the past $7-10$~Gyr.

As with all current studies of molecular gas at high redshift,
however, the results from PHIBSS rely on the uncertain conversion of
an observed molecular line luminosity (or some other indirect tracer)
to a measurement of the molecular gas mass.
Within local star-forming galaxies, molecular gas is primarily
composed of molecular hydrogen (\htwo{}) in giant molecular clouds
\citep[GMCs,][]{mckee2007}. Because it lacks a permanent dipole moment
to facilitate low energy dipole transitions, and the allowed
quadrupole transitions have very low probabilities
($\mathrm{A\sim10^{-7}~s^{-1}}$) and trace only warm
($\mathrm{T\sim5,000~K}$) gas, the \htwo{} in molecular clouds is
nearly invisible via direct emission.
Thus, the bright dipole transitions of the next most abundant
molecule, carbon monoxide (CO), are frequently used as molecular gas
tracers.
Extragalactic studies rely heavily on CO emission to determine the
molecular gas mass ($M_{\rm gas}$) from the observed
velocity-integrated CO luminosity (\lco{}) by way of the CO-H$_{2}$
conversion factor, \aco{}:
\begin{equation}
M_{\rm gas}=\alpha_{\rm CO}~\,~L'_{\rm CO}~,
\end{equation}
where $M_{\rm gas}$ (and thus \aco{}) includes a 36~per~cent
correction for Helium ($M_{\rm
  gas}=M_{\mathrm{H_2}}+M_{\mathrm{He}}=1.36~M_{\mathrm{H_2}}$).\footnote{Alternatively,
  the conversion factor (X$_{\rm CO}$) can be defined in terms of
  column density, such that $N_{{\rm H}_2}={\rm X_{CO}}I_{\rm CO}$.}
Although CO emission originates from only a fraction of the total
molecular gas, for self-gravitating GMCs, \aco{} is expected to be largely
independent of GMC mass, given that both the CO luminosity and
molecular gas mass at the CO emitting surface are proportional to the
enclosed total mass. In particular, the \aco{} value of a self-gravitating molecular cloud
is proportional to $n^{1/2}/T$ \citep{dickman1986}, 
where $n$ is the gas density and $T$ is the kinetic temperature.

Within the Milky Way, observations arrive at a value of $\alpha_{\rm
  CO,MW}=4.36$~\acounit{}~$\pm\sim30~{\rm per~cent}$ across a wide
range of GMC properties and environments \citep[e.g.][]{dickman1978,
  frerking1982, planck2011, ackermann2012}, indicative of a uniformity in
  the temperature and density of molecular clouds. 
As such, the canonical Milky~Way value for \aco{} is commonly utilized
as a constant conversion factor throughout a wide range of
extragalactic studies.
However, systematic variation of \aco{} with certain properties of the
interstellar medium (ISM, e.g.~metallicity) is both expected from
theoretical arguments and observed in a number of systems
\citep{leroy2011, sandstrom2013, strong2004, papadopoulos2012b}.

As with other molecular tracers, CO emission originates from gas that
is denser and further within molecular clouds than most of the
molecular hydrogen, resulting in a shell of \htwo{} between the CO
emitting surface and the edge of the molecular cloud. The size of
this shell (and thus the fraction of `CO dark' molecular gas) depends
on the efficiency of intervening dust at shielding CO from the
galactic radiation field. In this way, low-metallicity clouds with a
lower dust-to-gas ratio have higher \aco{} values.
The conversion factor is further coupled to the cloud structure
because CO emission is almost always optically thick
\citep{dickman1978, solomon1987}, and thus the CO intensity is
determined by the temperature and velocity dispersion at the
$\tau_{\rm CO}=1$ surface within the molecular cloud.
Clouds with sources of dispersion beyond self-gravity, such as tidal
disruptions or embedded stars, can have brighter
CO emission than self-gravitating molecular clouds of the same mass, thus 
lowering \aco{}.
Observationally, this effect primarily manifests itself as a
correlation between \aco{} and density --- molecular clouds in
high-density environments tend to be composed of `diffuse' molecular gas
containing embedded
stars that increase the velocity dispersion beyond that of a
self-gravitating cloud. Based on these arguments, \aco{} is commonly
parameterized as a function of the ISM density and temperature times a
separate function of metallicity \citep{bolatto2013, narayanan2012,
  tacconi2013}. As these ISM characteristics are known to vary with
redshift \citep[e.g.][]{magdis2012a, zahid2013, carilli2013}, \aco{}
very likely evolves with cosmic time.

Overall, observations indicate that the Milky Way value of
$\acom{}=4.36~\acounit{}$ is applicable for most galaxies, signifying
a relative uniformity in molecular cloud conditions
\citep{bolatto2008, leroy2011, sandstrom2013, schninnerer2010}. On the
other hand, observations within the Milky Way and in a relatively
small number of nearby and distant systems show variation in \aco{}
consistent with the expected dependence on ISM conditions described
above.
For example, variation in \aco{} with gas-phase metallicity has been
observed within the Local Group \citep{israel1997, leroy2011}, nearby
galaxies \citep{sandstrom2013}, and even a handful of systems at $z>1$
\citep{genzel2012}. While these results are all consistent with \aco{}
increasing by over an order of magnitude for metallicities less than
30~per~cent solar, measurements of \aco{} at a fixed metallicity have
a large dispersion, especially in medium-to-high metallicity
environments \citep{sandstrom2013}.

Variation in \aco{} with mass surface density has also been observed
in the Milky~Way, with $\gamma$-ray, dust, and dynamical observations
of the dense Galactic center arriving at a lower \aco{} than that
found in the solar neighborhood \citep{sodroski1995, oka2001,
  strong2004}.
Recent dust-based measurements of \aco{} within nearby galaxies find a
similar decrease in \aco{} at small galactic radii
\citep{sandstrom2013}. Moreover, for over a decade, it has been known
that the star-forming centers of Ultra-Luminous Infrared Galaxies
\citep[ULIRGs,][]{aaronson1984} require a lower \aco{} in order for
their CO luminosities to be consistent with their dynamical masses
\citep{downes1998}. As a result, a lower value of
$\acom{}=0.8$~\acounit{} is commonly adopted for mergers and ULIRGs
\citep{solomon1997, downes1998, tacconi2006, tacconi2008,
  krumholz2007, daddi2010}.

At high redshift, where molecular gas conditions may differ from those
locally, most measurements of \aco{} have been biased towards massive,
highly star-forming galaxies
that are not representative of the typical star-forming environment at
high $z$ \citep[e.g.][]{tacconi2008, magdis2011, spilker2015}. In the
new era of high-sensitivity radio telescopes, such as the NOrthern
Extended Millimeter Array (NOEMA) and the Atacama Large Millimeter
Array (ALMA), it is possible to probe molecular gas conditions at
$z\gtrsim2$ via CO \citep[e.g.][]{ivison2012, wang2013, tamura2014,
  messias2014, bethermin2016}. In order for these observations to
provide insight into the evolution of cosmic star-formation activity
since $z\gtrsim1$, a better understanding of \aco{} and its dependence
on galaxy properties for typical high-$z$ systems is critical.
Motivated by the dependence of \aco{} on ISM properties observed
locally, we study variations in \aco{} with integrated galaxy
properties, in particular mass surface density, in $38$ massive
star-forming galaxies at $z=1-1.5$ and a comparison sample of $164$
systems at $z\sim0.05$. In Section~\ref{sec:observations}, we detail
our observational datasets and galaxy
samples. Section~\ref{sec:inverseks} describes the technique that we
use for measuring \aco{}, while Section~\ref{sec:acodensity}
summarizes our results regarding the relationship between \aco{} and
mass surface density. Finally, Sections~\ref{sec:discussion} and
\ref{sec:conclusions} contain a discussion regarding the implications
of our results and as well as itemized conclusions. Throughout this
paper, we adopt a \cite{chabrier2003} initial mass function and a
$\Lambda$CDM cosmology with $H_0=70~{\rm km~s^{-1}~Mpc^{-1}}$,
$\Omega_M=0.3$, and $\Omega_\Lambda=0.7$.

\section{Data}
\label{sec:observations}
\subsection{High-\emph{z} Sample}
\begin{figure*}
\centering
\begin{tabular}{cc}
\hspace*{-0.175in}
\includegraphics[width=.5\linewidth]{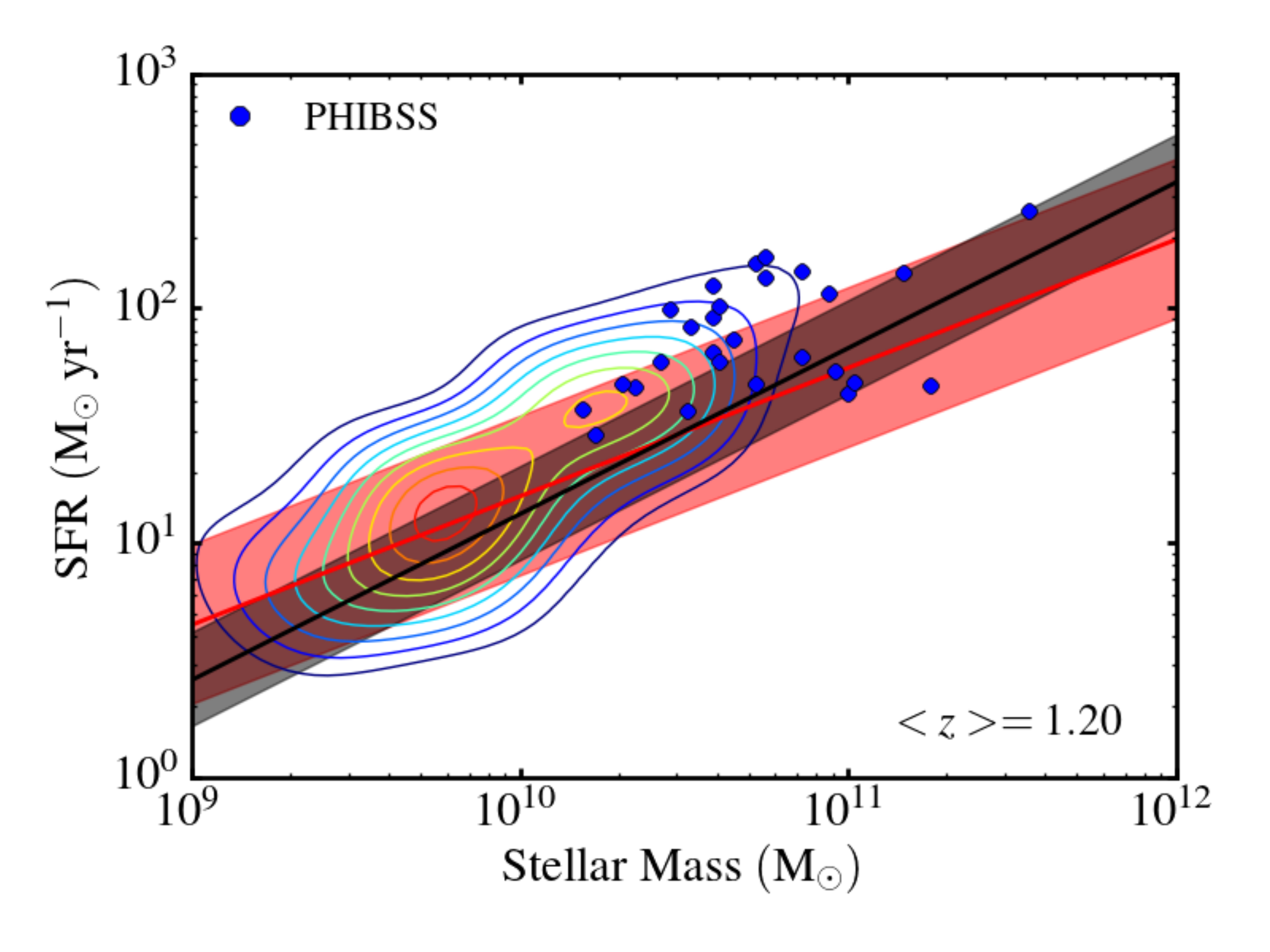}\label{fig:phibssms} & \includegraphics[width=.5\linewidth]{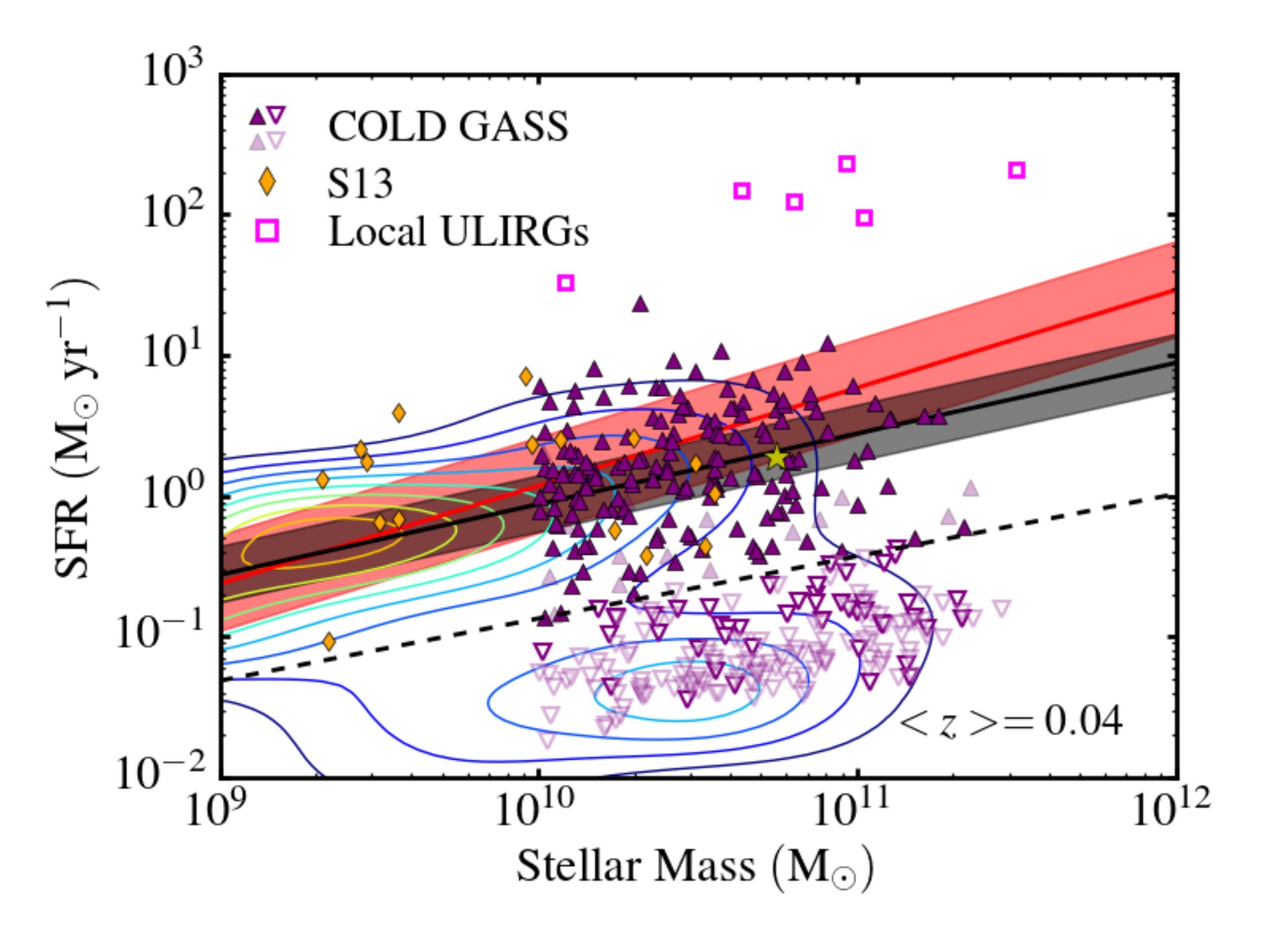}\label{fig:cgassms}
\end{tabular}
\caption{The location of the PHIBSS (\emph{left}) and COLD~GASS
  (\emph{right}) samples in the SFR-M$_*$ plane. Blue points in the 
  left panel correspond to the galaxies in our high-$z$ sample with
  star formation rates and stellar masses from the RAINBOW database
  \protect\citep{barro2011}. Purple triangles in the right panel
  correspond to galaxies in the low-$z$ sample with stellar masses and
  star formation rates from the MPA-JHU SDSS catalog.
  Downward open triangles represent galaxies below our quenching threshold and are not used
  in the analysis. Shaded points are not detected in CO also not used in the analysis.
  Additionally shown as orange diamonds are $17$ galaxies from
  \protect \cite{sandstrom2013} included in the low-$z$ sample. Stellar
  masses and star formation rates for these galaxies are taken from \protect
  \cite{kennicutt2011}, with stellar masses derived using $H$ band
  luminosities and SFRs derived from H$\alpha$+24\um{} luminosities.
  A comparison sample of local ULIRGs, taken from \protect \cite{downes1998}, 
  is shown with magenta squares, using stellar masses determined from the $H$ 
  band luminosity \protect \citep{zibetti2009} and SFRs calculated
  from the total infrared luminosity \protect \citep{kennicutt2012}. The
  location of the Milky~Way in this space is shown as the yellow star.
  The dashed line in the right panel indicates the threshold used to
  separate quenched and star-forming systems in our analysis, as
  defined in equation~\ref{eqn:quenchdef}.
  For comparison, the black and red lines illustrate fits to the
  star-forming ``main sequence'' at $z \sim 0.04$ and $z \sim 1.2$ from
  \protect\cite{speagle2014} and \protect \cite{whitaker2012},
  respectively, with the black and red shaded regions corresponding to
  the associated 1$\sigma$ scatter. To highlight observational
  scatter, contours of star-forming galaxies from publicly available
  catalogs are included. In the left panel, contours show galaxies
  from the RAINBOW database between $1<z<1.5$ with spectroscopic
  redshifts and stellar masses greater than $10^9~{\rm M}_\odot$, with
  contours every $75$ galaxies per square dex starting at $150$
  galaxies per square dex. In the right panel, contours show galaxies
  taken from MPA-JHU catalog with $0.025<z<0.05$ and stellar masses
  greater than $10^9~{\rm M}_\odot$, with contours every $3,000$
  galaxies per square dex. 
  Both of our samples draw from typical star-forming galaxies, as
  $56$~per~cent of the galaxies in the high-$z$ sample and
  $42$~per~cent of galaxies in the low-$z$ sample fall within
  $0.34$~dex of the main sequence from \protect
  \cite{speagle2014}.}
\label{fig:sfms}
\end{figure*}

Our high-$z$ galaxy sample includes $38$ star-forming systems at
$1<z<1.5$ from the IRAM Plateau de Bure HIgh-$z$ Blue Sequence Survey
\citep[PHIBSS;][]{tacconi2010, tacconi2013}. PHIBSS is one of the largest
high-redshift surveys of molecular gas to date, observing CO emission
from massive star-forming galaxies at $z=1.0-1.5$ and $z=2-2.5$ with
the IRAM Plateau de Bure interferometer. The PHIBSS sample at
$1<z<1.5$ is selected from the All-Wavelength Extended Growth Strip
International Survey \citep[AEGIS;][]{davis2007}, which provides
extensive multiwavelength imaging, including $V_{\rm F606W}$- and
$I_{\rm F814W}$-band observations with the \textit{Hubble Space
  Telescope} (\textit{HST}) Advanced Camera for Surveys (ACS) and
spectroscopic redshifts from the DEEP2 and DEEP3 Galaxy Redshift
Surveys for each galaxy \citep{newman2013, cooper2011, cooper2012,
  lotz2008}.
In addition, $J_{\rm F125W}$- and $H_{\rm F160W}$-band
\textit{HST}/WFC3-IR imaging from the Cosmic Assembly Near-infrared
Deep Extragalactic Legacy Survey \citep[CANDELS;][]{grogin2011,
  koekemoer2011,vanderwel2014} covers $32$ of the $38$ PHIBSS galaxies, while
\textit{HST}/WFC3-IR G141 grism observations exist for $28$ of the
$38$ systems in our high-$z$ sample (see Section~\ref{sec:grism}).

At $1<z<1.5$, PHIBSS targets `normal' star-forming galaxies, probing
down to star formation rates of $30~\mathrm{M_\odot~yr^{-1}}$ and a
stellar mass limit of $2.5\mscinot{10}~{\rm M_\odot}$.
This high-$z$ sample has a median stellar mass of
$6.7\times10^{10}~\msun$, within $0.2$~dex of $M^*$ at this redshift
\citep{ilbert2010}, and a median star formation rate of $87~\msun~{\rm
  yr}^{-1}$, corresponding to a sample predominantly composed of
Luminous Infrared Galaxies \citep[LIRGs;][]{sanders1996}.
As shown in Figure~\ref{fig:sfms}, all galaxies sit on or slightly
above the `star-forming main sequence', where the majority of star
formation occurs \citep[e.g.][]{noeske2007, rodighiero2011,speagle2014,
  whitaker2012}. Only $11$~per~cent of the galaxies are more than
$0.6$~dex above the main sequence, and the median difference between
the sample and the main~sequence \citep[taken from][]{whitaker2012} is
$0.3$~dex.

\begin{figure}
\centering
\includegraphics[width=1\linewidth]{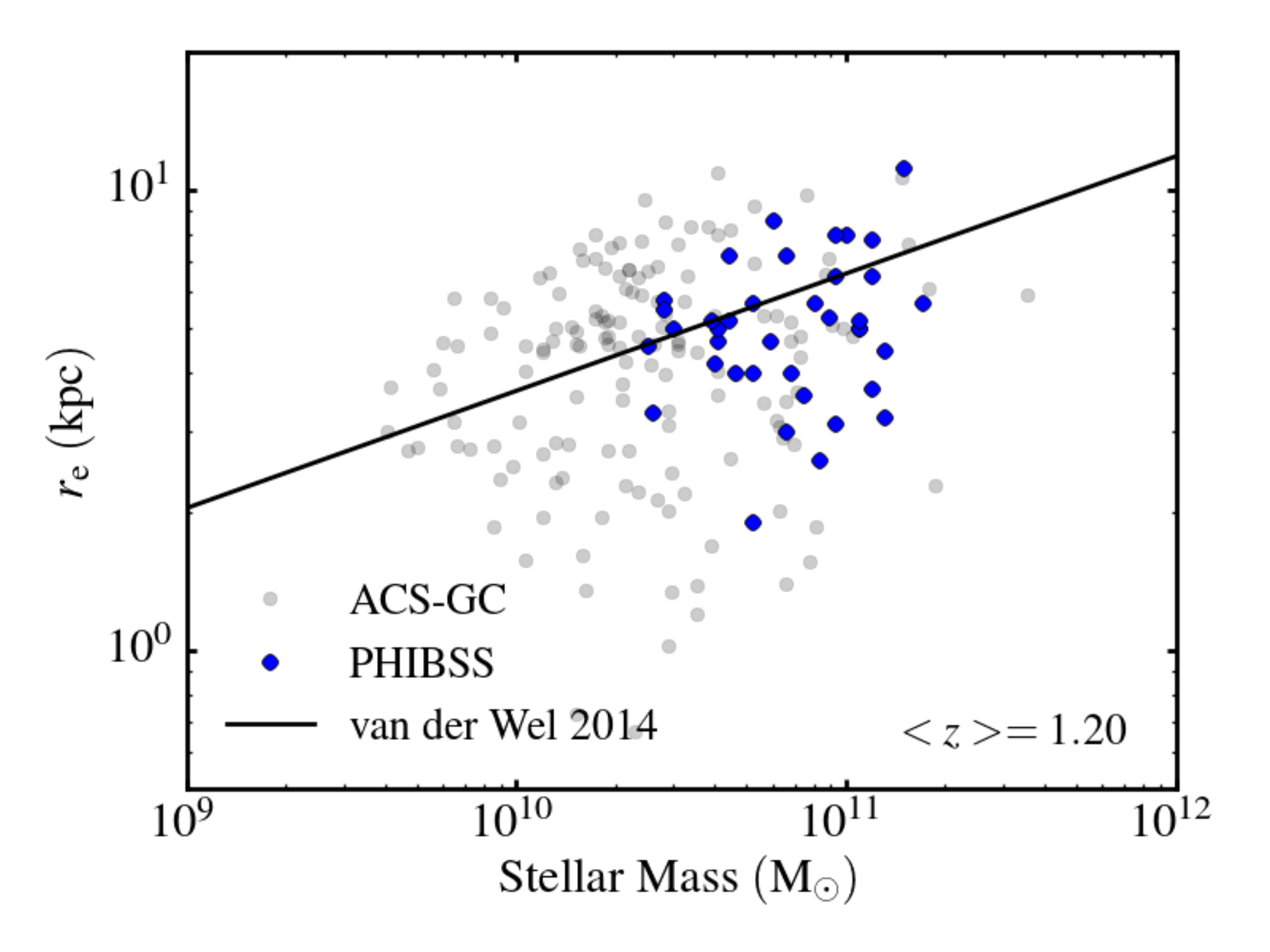}
\caption{
The size-stellar mass distribution for the PHBISS sample at $z \sim 1.2$
(blue circles). Here, we utilize the $I$-band (i.e.~rest-frame $B$-band)
size for each system, but find similar results when employing \ha{}
or rest-frame $V$- or $I$-band sizes (see Section \ref{sec:hasize}).
For comparison, gray points show the distribution of star-forming galaxies
at $1<z<1.5$, utilizing stellar masses from \protect\cite{barro2011} and 
\textit{I}-band sizes from the ACS-General~Catalog \protect\citep{griffith2012}.
In addition, the best fit size-mass relation for star-forming galaxies at $z\sim1$
from \protect\cite{vanderwel2014} is shown as the black line. The PHIBSS 
sample is not biased towards compact systems; instead, it is consistent with
being drawn from the parent population of massive star-forming galaxies at
intermediate redshift.}
\label{fig:sizemassbarrgriff}
\end{figure}

Although there is no size or morphological consideration in the
selection, we ensure that the size distribution of PHIBSS is
representative of typical high-$z$ star formers, utilizing a control
sample of star-forming galaxies at $1\le z\le1.5$ from the Advanced
Camera for Surveys\textendash{}General Catalog
\citep[ACS-GC;][]{griffith2012} and the RAINBOW database
\citep{barro2011}.\footnote{http://rainbowx.fis.ucm.es} The control
sample is restricted to galaxies with ${\rm
  SFR}>30~\mathrm{M_\odot~yr^{-1}}$, stellar mass uncertainty~$<~0.5$ dex and
$I$-band surface brightness profiles that can be reliably fit with a
one-component S\'{e}rsic profile (GALFIT\_FLAG=0). From this parent
sample (shown in Figure~\ref{fig:sizemassbarrgriff}), we build a
mass-matched distribution by randomly drawing (with replacement)
$1000$ galaxies with stellar masses within $0.3~{\rm dex}$ of each
PHIBSS object. Using this selection, $88$~per~cent of PHIBSS galaxies
are within the central two quartiles of the size distribution of the
control sample, indicating that our high-$z$ sample is not particularly
biased toward compact or diffuse systems.

Perhaps more importantly, it is unlikely that the PHIBSS sample is
biased towards abnormally gas-rich or gas-poor objects. While there is
no reference sample at $z\sim1$, to which to compare the observed gas
fractions, all of the PHIBSS \emph{targets} at $z \sim 1$ are detected in CO
emission and reside near the star-forming main sequence at this
redshift. Thus, there is no reason to believe that the
$10-80$~per~cent gas fractions detected in PHIBSS are not
representative.

\subsection{Low-\emph{z} Sample}

As a low-redshift comparison sample, with which to study potential
evolution in \aco{}, we utilize data from the CO Legacy Database for
GASS \citep[COLD~GASS;][]{saintonge2011a} along with published \aco{}
measurements from the literature \citep{sandstrom2013, leroy2011, 
papadopoulos2012}.
COLD~GASS is an extension of the \galex{} Arecibo SDSS Survey
\citep[GASS;][]{catinella2010}, which measured H{\scriptsize I} gas masses for
an unbiased, stellar mass-selected sample of galaxies at $0.025<z<0.05$
drawn from the Sloan Digital Sky Survey \citep[SDSS;][]{york2000}.
As the largest molecular gas survey in the local Universe, COLD GASS
provides an excellent comparison dataset at low redshift, probing a
stellar mass range comparable to that of our high-$z$ sample.

Similar to PHIBSS, COLD~GASS targets galaxies with stellar masses
ranging from $10^{10}-10^{11.5}~{\rm M_\odot}$, with a median value of
$10^{10.5}~{\rm M_{\odot}}$. Unlike PHIBSS, however, COLD~GASS targets
both quiescent and star-forming galaxies. To avoid including galaxies
with a potentially distinct star-formation process in our analysis, we
exclude quenched galaxies, defined as systems with
\begin{equation}
\log\left({\frac{\rm SFR}{\rm
      M_\odot~yr^{-1}}}\right)\le0.44~\log\left({\frac{M_*}{\rm
      M_\odot}}\right)-5.29 \, .
\label{eqn:quenchdef}
\end{equation}
This division is determined by a linear fit to
the minimum of the distribution of galaxies
at $z<0.05$ in the $\log{\rm sSFR}$-$\log{\rm M_*}$ plane
and is shown as the dashed line in
Fig.~\ref{fig:sfms}b. Further limiting our analysis to objects with a
CO detection only eliminates $15$ of the $179$ star-forming galaxies,
leaving $164$ in our final sample. These cuts generate a sample
similar to PHIBSS that is largely unbiased in terms of gas fraction
with an approximately uniform star-formation efficiency (assuming a Milky~Way 
\aco{}, the star-formation efficiency of our COLD~GASS sample has a scatter of
$0.34$~dex, consistent 	with the intrinsic scatter observed in local galaxies \citealt{leroy2013}). 
Finally, as with PHIBSS, the COLD~GASS sample is selected independent of galaxy
size or morphology, yielding a representative sampling of the
molecular gas in star-forming galaxies at $z\sim0$.

We combine the sample of integrated CO measurements from COLD~GASS
with spatially-resolved measurements of \aco{} for $17$ local galaxies
as published by \cite{sandstrom2013}. By minimizing the scatter in the
dust-to-gas ratio across $\sim$kpc-sized regions within nearby
($d<25~{\rm Mpc}$) galaxies, \cite{sandstrom2013} simultaneously
determine \aco{} and the dust-to-gas ratio, mapping their variation
within each galaxy. The sample is selected from the \textit{Spitzer}
Infrared Nearby Galaxies Survey \citep[SINGS;][]{kennicutt2003} and
the Key Insights into Nearby Galaxies: A Far-Infrared Survey with
\textit{Herschel} \citep[KINGFISH;][]{kennicutt2011} programs, with
follow-up $21$~cm observations as part of The H{\sc i} Nearby Galaxies
Survey \citep[THINGS;][]{walter2008} and CO $J$=($2\rightarrow1$)
observations from the HERA CO Line Emission Survey
\citep[HERACLES;][]{leroy2009}. Of the $26$ galaxies analyzed in
\citet{sandstrom2013}, the $9$ with inclination angles greater than
$65^\circ$ are excluded due to projection effects that complicate the
localization of the gas and dust emission. This sample has a median
stellar mass slightly less than that of the COLD~GASS and PHIBSS
samples (${<M_{*}>}=10^{9.8}~{\rm M_\odot}$), but represents the same
population of main-sequence galaxies as the rest of our low-$z$
sample, as shown in Figure~\ref{fig:sfms}.

Finally, our low-$z$ sample contains $10$ local ULIRGs with existing
\aco{} measurements from the literature. \citet{papadopoulos2012},
using observations of multiple CO and HCN transitions as constraints on large
velocity gradient modeling, determine \aco{} for a sample of $70$
local ULIRGs. The $10$ ULIRGs with independent dynamical masses
measured by \cite{downes1998} are included in our low-$z$
sample. These ULIRGs tend to be more massive and more intensely
star-forming than the rest of our low-$z$ sample (see
Fig.~\ref{fig:sfms}b), and provide a useful contrast to the more
typical galaxies in the \cite{sandstrom2013} and COLD~GASS samples.

\subsection{Galaxy Properties}
\label{sec:galaxyproperties}

For our high-$z$ sample, we employ the derived galaxy properties from
\cite{tacconi2013}, including ${\rm ^{12}CO}$~$J$=($3\rightarrow2$)
luminosities, star formation rates, stellar masses, and half-light
radii. For the COLD~GASS sample, we utilize ${\rm
  ^{12}CO}$~$J$=($1\rightarrow0$) luminosities from
\cite{saintonge2011a}, along with star formation rates and stellar
masses from the MPA-JHU catalog of derived galaxies properties for
SDSS~DR7 \citep{brinchmann2004,
  kauffmann2003},\footnote{http://wwwmpa.mpa-garching.mpg.de/SDSS/DR7/}
and \textit{u}- and \textit{r}-band Petrosian half-light radii
(PetroR50{\textunderscore}u, PetroR50{\textunderscore}r) from the
Sloan Digital Sky Survey Data Release~12 \citep{alam2015}. In this
section, we briefly describe the measurement of these and other galaxy
properties for the low-$z$ and high-$z$ samples.

\subsubsection{CO Luminosity (\lco{})}

For the PHIBSS galaxies, CO data cubes obtained with the IRAM Plateau
de Bure Millimeter Interferometer (PdBI) map the
$\mathrm{CO}$~$J$=($3\rightarrow2$) line at $\lambda_{\rm
  rest}=0.87~\mathrm{mm}$.
The CO spectra, obtained from observations
of the source in the most compact configuration, are fit to
single-component gaussians and integrated along the line of sight to yield the CO 
flux ($F_{\rm CO}$).
This is converted to a
CO~$J$=($1\rightarrow0$) luminosity using a constant
$L^\prime_{32}/L^\prime_{10}$ ratio of $0.5$, as suggested by CO ladder
observations of high-$z$ star forming galaxies
\citep{bauermeister2013, daddi2015, bolatto2015}.

In the COLD~GASS sample, each galaxy is targeted with a single
pointing of the IRAM $30$-m telescope to observe the
CO~$J$=($1\rightarrow0$) line at $\lambda_{\rm rest}=2.6~\mathrm{mm}$
\citep{saintonge2011a}. For targets larger than the $22$ arcsec beam,
aperture corrections are applied by scaling the total flux by the
$z$-band radius measured from the SDSS photometry; however, because
COLD~GASS galaxies are selected to be at $z>0.025$, only a small
number of systems require this correction. The aperture-corrected CO
flux is then integrated along the velocity width of the line and
converted to a CO luminosity.

\subsubsection{Stellar Masses, Star Formation Rates, and Effective Radii}
\label{sec:opticalgalaxyproperties}

Stellar masses in both the high-$z$ and low-$z$ samples are determined
by fitting template spectral energy distributions (SEDs) to observed
photometry. For the high-$z$ sample, the SEDs are fit to $26$-band
AEGIS photometry, yielding a typical uncertainty of about
$0.13$~dex. For the low-$z$ sample, the SEDs are fit to SDSS
$ugriz$-band photometry corrected for emission-line contributions,
yielding a typical uncertainty of approximately $0.1$~dex. Further
details regarding the SED fitting can be found in \cite{tacconi2013}
and \cite{kauffmann2003}.

Following \cite{wuyts2011}, star formation rates in the high-$z$
sample are calculated according to the calibrated relation from
\citet{kennicutt1998}:
\begin{equation}
{\rm SFR}=1.087\mscinot{-10}~[L_{\rm IR}+3.3\nu L_\nu(2800{\rm
  \AA})]~\msun~{\rm yr}^{-1}, 
\end{equation}
which accounts for both unobscured and obscured star formation as
traced by the observed UV and IR luminosities, respectively.
The rest-frame UV luminosity is calculated from the best-fit SED,
which is primarily constrained by CFHT $g^\prime$- and $i^\prime$-band
photometric observations that sample rest-frame $2800$\AA~at $z \sim
1-1.5$.
The total infrared luminosity, tracing obscured star formation and representing
$\sim80\%$ of the total star-forming activity, is
calculated from a single far-IR SED extrapolated from the
\textit{Spitzer} MIPS $24\mu$m emission. This extrapolation is
supported by {\it Herschel} PACS observations of star-forming galaxies
from $z=0-2.5$ \citep{elbaz2011}, and is sensible for all but the most
extreme star-forming systems. Star formation rates for the COLD~GASS
sample are determined from emission-line luminosities, corrected for
aperture losses based on SED fits to $ugriz$ photometry. Further
details regarding the SFR measurements can be found in
\cite{brinchmann2004}.

We use effective radii (\re{}) from \cite{tacconi2013} and the SDSS
DR12 \citep{alam2015} as size measurements for our high-$z$ and
low-$z$ samples, respectively.
Unless otherwise stated, in our high-$z$ sample, effective radii are
derived from a single component S\'{e}rsic fit \citep{sersic1968}:
\begin{equation}
I(r)\propto \exp[{\big( \frac{r}{r_e}\big)^{1/n}-1}] 
\end{equation}
to single-orbit \textit{HST}/ACS F814W (rest-frame $B$-band) images.

As illustrated in Fig.~\ref{fig:sizemassbarrgriff}, the resulting
effective radii in the high-$z$ sample are representative of the star-forming population at
this redshift. Moreover, the rest-frame $B$-band sizes are consistent
with the measured extent of the \ha{} emission, as shown in
Section~\ref{sec:hasize}.
While F606W (rest-frame NUV) imaging is 
available for only $23$ of our high-$z$ objects, our results are
qualitatively unchanged when using sizes derived from {\it HST}/ACS
F606W (versus F814W) imaging.
For the low-$z$ sample, we utilize the Petrosian half-light radius
(PetroR50{\textunderscore}u and PetroR50{\textunderscore}r) as
measured from the SDSS $u$- and $r$-band imaging.

\subsection {\emph{HST}/WFC3-IR Grism Observations}
\label{sec:grism}

A subset ($28$) of the galaxies in our high-$z$ sample were
observed with the {\it HST}/WFC3-IR G141 grism as part of a Cycle~19
GO program (PID 12547, PI: Cooper) or as part of the \dhst{} program
\citep[PID 12177, PI: van Dokkum;][]{brammer2012, vandokkum2013}.
The spectral range of the G141 grism extends from $1.1$\um{} to
$1.65$\um{}, which traces \ha{} emission at $0.7 \lesssim z \lesssim
1.5$. Both the GO and \dhst{} observations are two orbits deep,
corresponding to an approximate $5\sigma$ \ha{} detection limit of
${\rm\sim5\times10^{-17}~erg~s^{-1}~cm^{-2}}$. Accompanying direct image
exposures provide the necessary information for source identification
and contamination corrections. The WFC3-IR images were processed
utilizing the {\sc calwf3} reduction pipeline, with the grism spectra
and corresponding contamination images extracted using the {\sc aXe}
reduction pipeline (\citealt{kummel2009}; Weiner et al.~in prep).
All but three (i.e.~$25$ out of the $28$) galaxies are detected in
\ha{} emission at the $5\sigma$ level; we exclude an additional two
systems 
for which the \ha{} emission is located too close to the edge of the
WFC3-IR detector.

\begin{figure}
\centering
\includegraphics[width=1\linewidth]{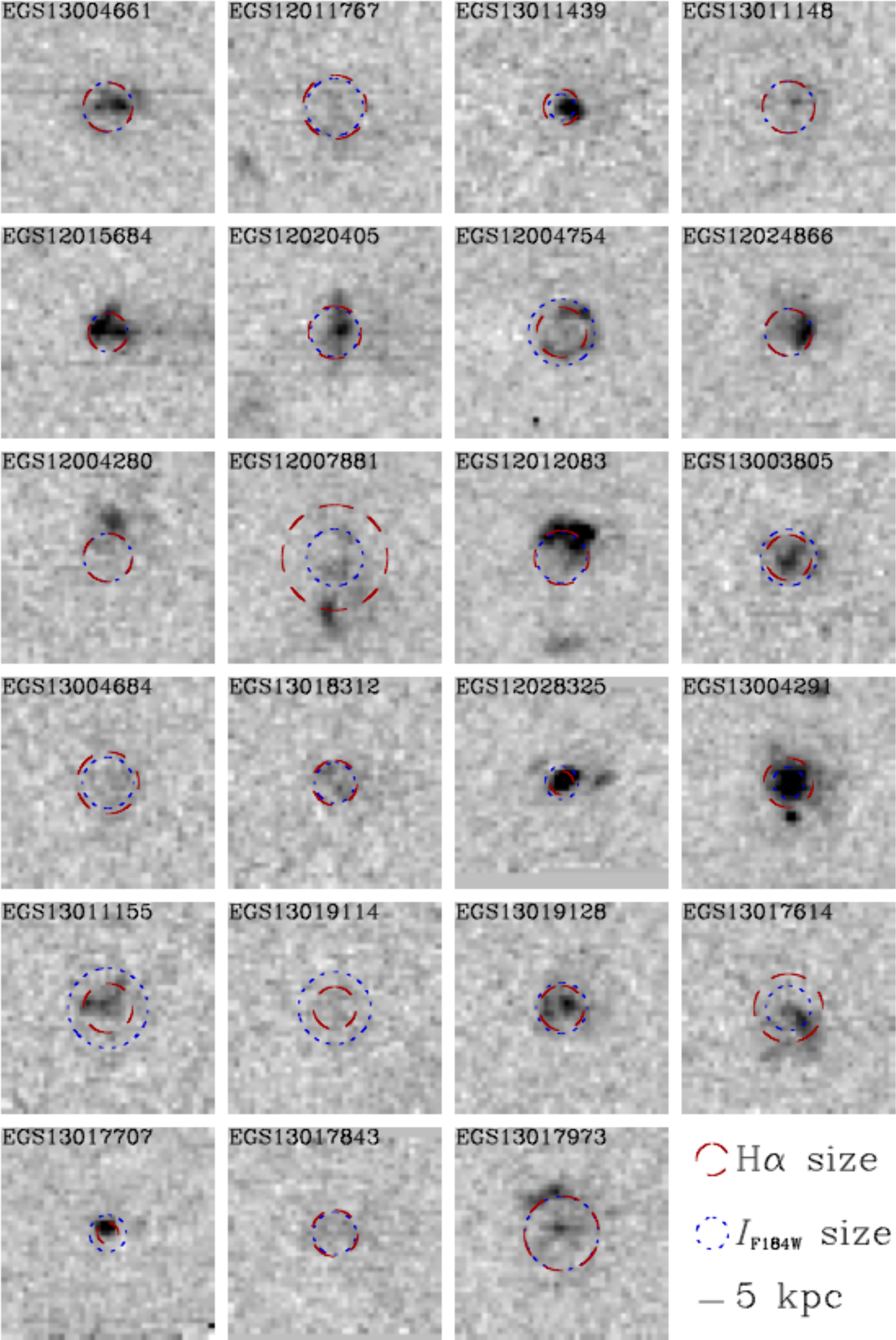}
\caption{ \ha{} images derived from \textit{HST}/WFC3-IR G141 grism
  observations for a subset of our PHIBSS sample at $z \sim 1$. For
  each galaxy, the measured \ha{} size and {\it HST}/ACS
  \textit{I}-band size are illustrated by the dashed red circle and
  dot-dashed blue circle, respectively. In general, the \ha{} and
  broad-band sizes are in very good agreement, with an average
  difference of less than $0.1$~physical~kpc. As a reference, the
  legend includes a bar corresponding to $5$~kpc (physical) at the
  redshift of our high-$z$ sample ($z=1.2$).}
\label{fig:justhaimages}
\end{figure}

\subsubsection{\ha{} SFR}
\label{sec:hasfr}

The H$\alpha$ emission probed by the {\it HST}/WFC3 grism observations
provides a valuable cross-check of the SFR measurements inferred from
broad-band photometry.
The advantage of using \ha{} emission as an independent SFR indicator
is two-fold. First, \ha{} emission from H{\scriptsize II} regions
around young ($<10$~Myr old) stars provides a more instantaneous
star-formation indicator than UV and IR broad-band emission. Second,
SFR measurements from \ha{} are free from the assumptions and
uncertainties of SED fitting techniques. We calculate the star
formation rate ($\mathrm{SFR_{H\alpha}}$) from the observed \ha{}
luminosity ($L_{\rm H\alpha}$) according to the empirical relation of
\cite{kennicutt2012} adjusted to a Chabrier IMF following
\citet{twite2012}:
\begin{equation}
{\mathrm{SFR_{H\alpha}}=4.6\times10^{-42}~L_{\rm
    H\alpha}\times10^{0.4A_{\rm H\alpha}}}~, 
\end{equation}
where the \ha{} luminosity is determined by integrating the \ha{}
emission from the continuum subtracted one-dimensional grism
spectrum. Continuum emission is fit to a third-order polynomial, with
a $\sim400$\AA\ region around the emission line masked from the fit.

To correct for contamination from nearby [N{\scriptsize{II}}]
$\lambda\lambda6548,6583$ emission, we use the strong correlation
between the [N{\scriptsize{II}}]--\ha{} ratio and gas-phase
metallicity \citep{pettini2004}. The metallicity of each system is
inferred from the measured stellar mass according to the
mass-metallicity relation of \cite{zahid2014} and then converted to
the [N{\scriptsize{II}}]/\ha{}-based metallicity scale of
\cite{pettini2004} using the conversion from \cite{kewley2008}:
\begin{equation}
\mathrm{12+\log{O/H}=8.90+0.57\times[N{\textsc{ii}}]/H\alpha}~.
\end{equation}
The N{\scriptsize{II}} emission (typically $30$~per~cent of the
observed emission line) is then subtracted from the integrated
luminosity to determine the \ha{} luminosity.
We estimate the relevant
extinction (${A_{\rm H\alpha}}$) from the observed infrared excess
(IRX):
\begin{equation}
  A_{\rm H\alpha}=1.37\log[{1+3.89L_{\rm 24 \mu m}/L_{\rm FUV}}]~,
\end{equation}
where ${{\rm IRX}=\log[{L{\rm (TIR)}/L{\rm(FUV)}_{\rm obs}}}]$
\citep{hao2011}.
To calculate the IRX, we use a combination of rest-frame $24$\um{} and
FUV synthetic photometry from SED fits available in the RAINBOW
database.

The resulting \ha{} star formation rates agree with
the corresponding UV+IR star formation rates from \cite{tacconi2013},
with the \ha{} SFRs only $0.03$~dex lower on average.
The scatter between the \ha{} and UV+IR star formation rates, however,
is large, with an rms difference of $0.49$~dex. While some of this scatter can be attributed to the 
	uncertainty in the [N{\scriptsize{II}}] subtraction  ($\sim0.1$~dex) and intrinsic scatter in the IRX-based 
	${A_{\rm H\alpha}}$
estimates ($\sim0.13$~dex), the large scatter is likely
due to $73$~per~cent of the sample having an \ha{} extinction greater
than $2.5$~mag, beyond the calibration of \cite{hao2011}. As a result,
we employ the UV+IR star formation rates for the remainder of the
analysis, though our results are qualitatively unchanged if we use the
\ha{} star formation rates.

\subsubsection{\ha{} Size}
\label{sec:hasize}

It is known that galaxy sizes vary with rest-frame color, with red
images tracing the more concentrated stellar component and blue images
tracing the extended star-forming component \citep{nelson2012,
  vanderwel2014}. In light of this, we test whether the $I$-band radii
are an accurate indicator of the size of the star-forming component in
the $z\sim1$ galaxies by measuring effective radii from the grism
\ha{} images. Because of its low spectral resolution, 
the {\it HST}/WFC3-IR G141 grism spectra can be thought of as a
sequence of adjacent images, taken at 101\AA{} increments, such that
any observed structure along the spectral direction must be spatially
extended as outflows would need a radial velocity of $V>{\rm
  2000~km~s^{-1}}$ to yield a resolved Doppler shift. For each row in
the grism image, we mask the \ha{} emission and fit a third-order
polynomial to the continuum, using the same procedure as outlined in
Section~\ref{sec:hasfr}. This continuum is then subtracted from the
contamination-subtracted grism image to generate maps of the \ha{}
emission (see Fig.~\ref{fig:justhaimages}). The advantage of the grism
is apparent; spatial structure that would have been missed by a
high-resolution spectroscopic observation is readily apparent.

Due to strong asymmetry and clumpiness in the H$\alpha$ emission,
measuring a size (i.e.~$r_e$) via a parametric fit to the light profile
is often problematic. Instead, we measure \re{} by constructing
curves of growth from the \ha{} images. The curves of growth are
centered on the wavelength of \ha{} in the spectral direction (as
determined from the DEEP2/DEEP3 spectroscopic redshift) and the center
of the continuum in the spatial direction, which is taken to be the
median centroid of Gaussian fits to each column of continuum
emission. Rather than using a fixed aperture to determine the total
enclosed flux, we utilize the first aperture, for which the curve of
growth decreases in two subsequent apertures (spaced at radial
increments of $0.13^{\prime\prime}$).
The radius at which the curve of growth passes half of this total flux
is identified as the effective radius.

As illustrated in Figure~\ref{fig:justhaimages}, the measured \ha{}
sizes agree with those derived from the {\it HST}/ACS $I$-band
imaging; the median absolute difference between the two sizes
($|r_{\rm H\alpha}-r_I|$) is $0.07$~kpc with an rms scatter of
$1.6$~kpc.
While individual galaxies may deviate by greater than the statistical
uncertainty of the size measurement (typical uncertainty is $0.14$~kpc
or roughly $3$~per~cent), they are within the systematic uncertainty
associated with non-uniform extinction and ellipticity. The agreement
in size measurements is consistent with both the \ha{} and UV
emission being dominated by star formation. \cite{nelson2012} employ
a similar procedure to compare \ha{} sizes of $z\sim1$ galaxies on the
main sequence with rest-frame $R$-band sizes and find that \ha{}
emission tends to be more extended than $R$-band emission in
star-forming galaxies, especially for the most extended ($r_{e}>5$~kpc
$R$-band) systems.
Although $9$ galaxies in our high-$z$ sample with grism observations
have sizes greater than $5$~kpc, we do not see strong evidence of this
bias. This is likely because the $I$-band measurements are blueward of
the rest-frame $4000{\rm \AA{}}$ break for most of our high-$z$ sample
and primarily trace the star-forming component. Among systems with $H_{\rm F160W}$
observations, our results are consistent with those of \cite{nelson2012}, with ${\rm H}\alpha$ sizes
$17\%$ larger than $H_{\rm F160W}$ (rest frame $R$ band) sizes on average.
As we have no evidence
for bias in the $I$-band size measurement, and the $I$-band ACS images
have higher spatial resolution, we use the $I$-band sizes throughout
the remainder of our analysis.

\section{Estimating the \boldmath CO-H$_{2}$ Conversion Factor}
\label{sec:inverseks}

In order to constrain \aco{}, we require an independent measurement of
the molecular gas mass -- i.e.~one not derived from the observed CO
line flux. Often, studies of \aco{} utilize \htwo{} masses inferred
from the dynamics of resolved gas clouds \citep{planck2011}, other
cold molecular gas tracers
\citep[e.g.~$^{13}$CO,][]{papadopoulos2012}, or cold dust emission
coincident with molecular gas \citep{leroy2011, sandstrom2013}. For
our samples, however, we lack the sensitive complementary observations
necessary for these traditional techniques.\footnote{Upcoming observations
as part of PHIBSS2 (Tacconi et al.~in prep) will resolve CO emission within
a sample of galaxies at $z\sim1$, providing additional constraints on our measurements.}

While we are unable to apply these conventional methods to measure
$M_{\rm gas}$, we can indirectly probe the molecular gas reservoir via
the observed star formation rate, assuming a particular relationship
between the current star-forming activity and the total molecular gas
mass \citep[e.g.][]{genzel2012}. This relationship is quantified in
terms of the molecular gas depletion time, $\tdep{}=M_{\rm gas}/{\rm
  SFR}$, which is shown to be largely independent of atomic and
molecular gas fractions at $z \sim 0$
\citep{bigiel2008,bigiel2011}. Furthermore, millimeter observations of
molecular gas in local main-sequence galaxies measure an approximately
 constant depletion time on kpc-scales, mostly
independent of gas surface density and indicative of a homogeneous
star-formation process \citep{bigiel2011, leroy2013, huang2014}.

\begin{figure}
\centering
\includegraphics[width=1\linewidth]{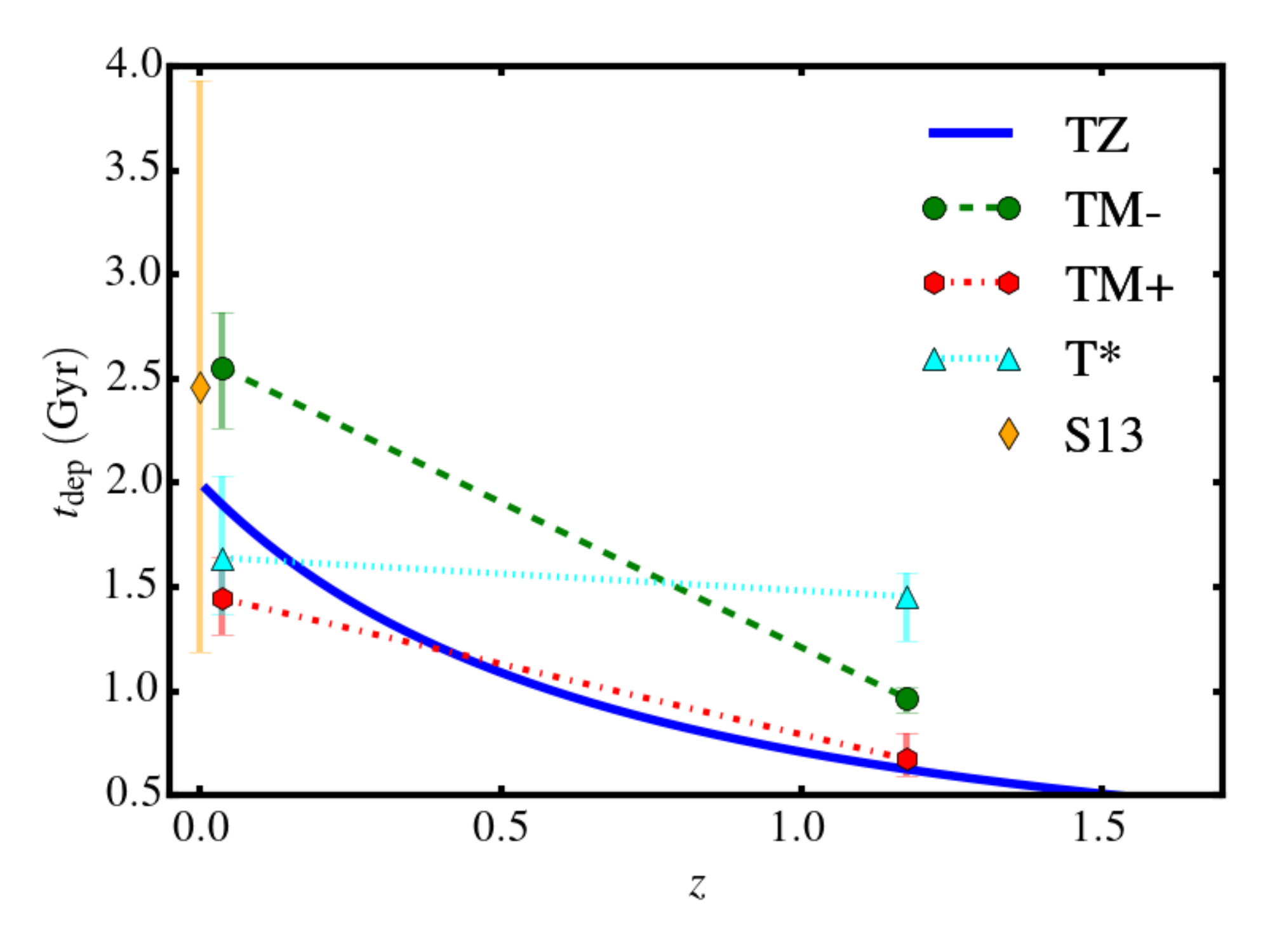}
\caption{A summary of the depletion time models adopted in our
  calculation of \aco{}.
  As model \ref{tdepz} is solely dependent on
  $z$, it is shown as a solid line from $z\sim0$ to $z\sim1$. For
  models~\ref{tdepdave}, \ref{tdepzmpos}, and \ref{tdepssfr}, points
  represent the median depletion time value within the COLD~GASS and
  PHIBSS samples and error bars correspond to the $33-66$ percentile
  range of the corresponding \tdep{} distribution.
  For context, the orange point illustrates the median, CO-independent, \tdep{} 
  measurement within the \protect \cite{sandstrom2013} sample, with error bars
  again corresponding to the $33-66$ percentile
  range of the distribution. The difference in
  the assumed dependence of \tdep{} on $z$ and stellar mass across the
  four models allows us to probe potential systematic uncertainties
  associated with the relationship between \aco{} and total mass
  surface density (or metallicity). Although models~\ref{tdepdave},
  \ref{tdepzmpos}, and \ref{tdepssfr} are all normalized to $2$~Gyr at a stellar mass
  of $10^{10.7}~{\rm M_\odot}$ at $z=0$, the different stellar mass dependencies of the models
  result in different median depletion times in the low-$z$ sample.}
\label{fig:tdepmodels}
\end{figure}

Assuming a universal depletion timescale, we can determine the
molecular gas mass from a galaxy's star formation rate, leading us to
an expression for \aco{}:
\begin{equation}
\label{eqn:acoeqn}
\alpha_{\rm CO}=\frac{{\rm SFR}~\;t_{\rm dep}}{L^\prime_{\rm CO}} \, .
\end{equation}
At present, there is considerable uncertainty regarding the dependence
of \tdep{} on galaxy properties and cosmic time. However, observations
of nearby galaxies find evidence for variation in \tdep{} with stellar
mass \citep{saintonge2011b, leroy2013, boselli2014} and specific
star-formation rate \citep{saintonge2011b, saintonge2012,
  boselli2014,genzel2015}. In addition, reduced depletion times have been observed within dense
  galaxy centers \citep{leroy2013}, and trends between \tdep{} and gas density
  \citep{kennicutt1998,kennicutt2012,genzel2010}, stellar surface density \citep{boselli2014}, and 
  the dynamical time of the gas \citep{kennicutt1998,kennicutt2007} have been observed.
To ensure that any observed relationships between \aco{} and galaxy
properties within our samples are not simply artifacts of an
underlying correlation with depletion time, we use four
physically-motivated models for \tdep{}, covering a range of possible
relationships between \tdep{} and galaxy properties\footnote{We do not include a density-dependent 
	depletion time model in our primary analysis, but our results remain qualitatively unchanged when a density-dependent 
	\tdep{} 
	law is adopted (see Section~\ref{sec:tdepsigma}).}
	and/or redshift.
The depletion time models we employ are: 
\begin{align*}
\tdep{}~{\rm (Gyr)}&=2\times (1+z)^{-1.5}, \tag{TZ}\label{tdepz} \\[1em]
\tdep{}~{\rm (Gyr)}&=2\times (1+z)^{-1.5}\left(\frac{M_*}{10^{10.7}~{\rm M}_\odot}\right)^{0.36}, \tag{TM+}\label{tdepzmpos} \\[1em]
\tdep{}~{\rm (Gyr)}&=2\times \left(\frac{t_H}{13.5~{\rm Gyr}}\right) \left(\frac{M_*}{10^{10.7}~{\rm M}_\odot}\right)^{-0.3}, \tag{TM-}\label{tdepdave}\\[1em]
\tdep{}~{\rm (Gyr)}&=2\times \left(\frac{\rm sSFR}{\langle {\rm sSFR}
    \rangle_z}\right)^{-0.44} \, , \tag{T*}\label{tdepssfr}
\end{align*}
where $M_{*}$ and sSFR are the stellar mass and specific star
formation rate of the system, $t_{H}$ is the Hubble time at the
appropriate redshift, and $\langle~{\rm sSFR}~\rangle_z$
is a redshift-dependent normalization parameter.

Model~\ref{tdepz}, which is independent of galaxy properties and
smoothly decreasing with increasing redshift, represents the simplest
of our assumed \tdep{} models. The normalization is set by COLD~GASS
observations at $z\sim0.1$ \citep{saintonge2011b}, while the redshift
dependence, which is slightly steeper than that derived in
\cite{tacconi2013} or \cite{genzel2015}, traces the evolution of the
dynamical time and reproduces complementary observations at higher $z$
\citep{bauermeister2013b, geach2011, saintonge2013, magdis2012b}.
The positive correlation between \tdep{} and $M_{*}$ in
model~\ref{tdepzmpos} is taken from the \cite{saintonge2011b} fit to
COLD~GASS observations, and is interpreted as the consequence of bulge
growth stabilizing gas in the disk and suppressing star formation
\citep[e.g.][]{martig2009}.

The mass and redshift dependence in model~\ref{tdepdave} matches the
prescription of \cite{dave2012}, found by associating the depletion
time with the local dynamical time.
\footnote{\cite{dave2012} do not
  distinguish between atomic and molecular gas; however, the molecular
  component is expected to be dominant with our sample, such that the
  inclusion or exclusion of atomic gas should not significantly affect
  our results.}
This model is normalized such that galaxies at $z=0$ with a stellar
mass of $10^{10.7}~{\rm M_\odot}$ have a $2$~Gyr depletion time,
consistent with models~\ref{tdepz} and \ref{tdepzmpos}. Finally,
model~\ref{tdepssfr} replicates the measured dependence of \tdep{} on
sSFR as observed by \cite{saintonge2011b}, normalized to $2$~Gyr at a
sSFR of $10^{-10.4}~{\rm yr^{-1}}$
for the low-$z$ sample and $10^{-9.2}~{\rm yr^{-1}}$ for the high-$z$
sample.\footnote{The redshift dependence of this normalization is 
	chosen to match the global sSFR evolution from \cite{gonzalez2008}}
 This sSFR dependence is consistent with the
findings of \cite{genzel2015}, who, using a broad range of CO- and
dust-based molecular gas measurements, report $\tdep\propto({\rm
  sSFR}/{\rm sSFR_{MS}})^{-0.49}$, where ${\rm sSFR_{MS}}$ is the sSFR
of a galaxy on the main sequence at a given stellar mass and redshift
(see also Tacconi et al.,~in prep).
This model attempts to incorporate variation in star-formation
efficiency that may drive the scatter in the star-forming main
sequence. Galaxies with lower depletion times necessarily have a
higher sSFR, as they are able to convert their molecular gas reservoir
into star formation more efficiently.

Figure~\ref{fig:tdepmodels} illustrates the median redshift dependence
(and scatter) of \tdep{} values for each model as applied to the
COLD~GASS and PHIBSS samples.
For our four models, the evolution in the resulting depletion times,
as well as their scatter at fixed $z$, varies significantly. The
difference in assumed dependence on $z$ and stellar mass, in
particular, allows us to probe potential systematic uncertainties
associated with the relationship between \aco{} and total mass surface
density.
Model~\ref{tdepdave} yields the strongest evolution in \tdep{} due, in
large part, to the difference in median stellar mass between our
low-$z$ and high-$z$ samples and the assumed mass dependence of that
model.
Altogether, it is important to note that our approach for estimating
\aco{} will not be accurate on an object-by-object basis. The inferred
\aco{} estimates are meaningful on average, however, allowing us to
explore -- in a statistical sense -- the global trends of \aco{} with
galaxy properties and/or redshift.

\begin{figure*}
\centering
\includegraphics[width=1\linewidth]{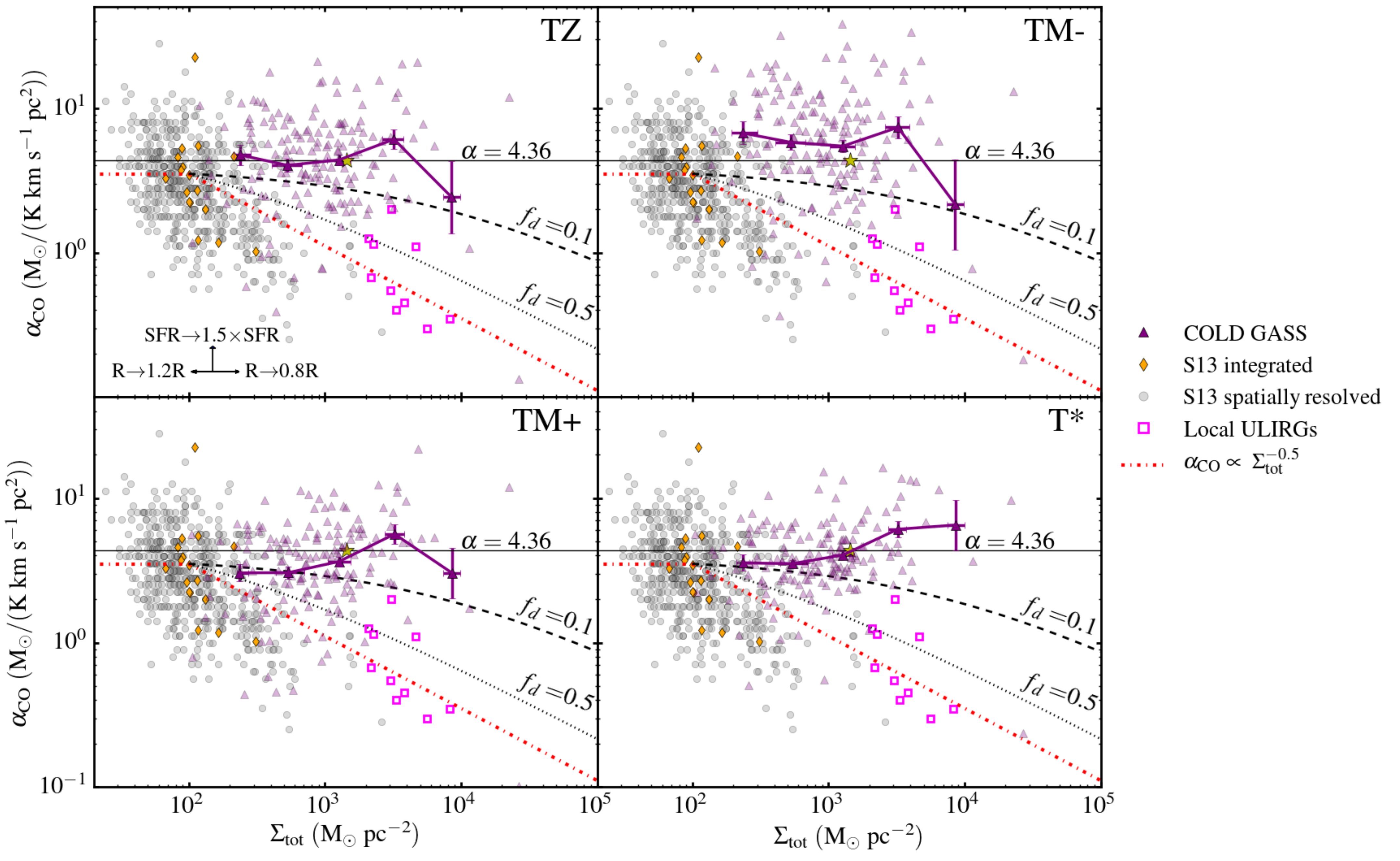}
\caption{The dependence of \aco{} on total mass surface density
  (\sigtot{}) in the low-$z$ sample. The four panels, as labeled,
  correspond to \aco{} values inferred via each of our assumed \tdep{}
  models. Across all panels, grey points show spatially-resolved,
  dust-based measurements of \aco{} in local galaxies from
  \protect\cite{sandstrom2013}, exhibiting both regions diffuse (high \sigtot{} and low \aco{}) and self-gravitating (Milky~Way \aco{}) gas,
    with orange diamonds representing the
  galaxy-averaged values.
  Additionally, magenta points correspond to
  measurements of local ULIRGs, with \sigtot{} from \protect
  \cite{downes1998} and \aco{} based on single component fits from
  \protect \cite{papadopoulos2012}. The {\it faint} purple triangles
  show our measurements for the $164$ star-forming galaxies in our
  low-$z$ sample from COLD GASS, with the {\it dark} purple points
  illustrating the mean \aco{} for this sample across $5$ distinct
  bins in \sigtot{} ranging from $125~{\rm M_\odot~pc^{-2}}<\sigtot{}<1.6\times10^4~{\rm M_\odot~pc^{-2}}$.
   For the binned measurements, the horizontal error
  bar indicates the $25-75\%$ range in \sigtot{} for the population
  within the given bin, and the vertical error bar shows the $1\sigma$
  error on the mean \aco{} value. Vectors in the lower-left corner
  of the top-left panel indicate how a change in SFR or \re{} affect
  our measurements of \aco{} and \sigtot{}. The red, dot-dashed line
  indicates the relationship between \aco{} and \sigtot{} for molecular gas
  within galaxies, as suggested by
  \protect\citet{bolatto2013}, Eq.~31, shifted to the average metallicity of
  our low-$z$ sample according to the mass-metallicity relation of \citet{zahid2014}.
   The dotted and dashed black lines denote models for the
  dependence of \aco{} on surface density assuming a mix of diffuse
  molecular gas (with $\aco{}\propto\Sigma_{\rm tot}^{-0.5}$) and gas
  in Milky~Way-like GMCs (with a constant \aco{}; see equation
  \ref{eqn:avgeqn}). The dotted line represents a $50-50$ mix, by mass, of
  diffuse gas and GMC-like gas, and the dashed line represents a mix
  of $10\%$ diffuse gas and $90\%$ GMC-like gas.
    	 The \aco{} points
  within the COLD~GASS sample do not show a significant decrease with
  increasing total mass surface density, consistent with a population
  that contains a small fraction molecular gas in diffuse, extended
  clouds.}
\label{fig:acodensity4plotcgas}
\end{figure*}

\section{Dependence of $\mathbf{\aco{}}$ on \boldmath${\Sigma_{\rm
      \lowercase{tot}}}$} 
\label{sec:acodensity}

The expected relationship between \aco{} and total (baryonic) mass
surface density (\sigtot{}) in our sample depends on the primary phase
of molecular gas within the systems. The \aco{} value of GMCs in the
Milky~Way disk is largely insensitive to the surrounding galactic
environment, with no significant evidence for a correlation between
\aco{} and \sigtot{} \citep{solomon1987, planck2011}. A correlation
between \aco{} and \sigtot{} has been observed, however, within
`diffuse' clouds of molecular gas \citep{downes1998, sandstrom2013}.
For example, at the centers of local ULIRGs, observations indicate that
molecular gas primarily exists as an extended molecular cloud (100s of
pc in diameter) containing stars, individual molecular clouds, and
diffuse molecular gas \citep{downes1998,leroy2015}. This diffuse
gas has a higher velocity dispersion (and thus lower \aco{}) than a
similarly massive, but isolated cloud due to the contribution of
enclosed stars to the gravitational potential. For such a cloud, the
dependence of \aco{} on total mass surface density
roughly follows the relation: $\aco{}\propto\Sigma_{\rm tot}^{-0.5}$
\citep{bolatto2013}, reflecting the dependence of the velocity
dispersion (i.e.~$L^\prime_{\rm CO}$) on total mass surface
density \citep{downes1998}. This description is consistent with
observations of local ULIRGs \citep{papadopoulos2012}, simulations of 
molecular gas in disks and
mergers \citep{narayanan2011}, as well as (on a smaller scale) the
lower \aco{} values observed at the centers of local star-forming
systems \citep{sandstrom2013}.

While this correlation between \aco{} and \sigtot{} is only expected to lower \aco{} in a diffuse
molecular cloud within a high-density
region of a galaxy, it may dictate the galaxy-wide \aco{} value if a
significant fraction of the total molecular gas is in this phase.  To
first order, the galaxy-integrated \aco{} can be thought of as the
mass-weighted average of \aco{} in diffuse
cloud(s)
($\alpha_D$) and in the rest of the galaxy ($\alpha_G$), assumed to be
composed of self-gravitating GMCs:
\begin{equation}
\label{eqn:avgeqn}
\langle\aco{}\rangle = \frac{\alpha_D\alpha_G}{f_d\alpha_G+(1-f_d)\alpha_D}
\, ,
\end{equation}
where $f_d$ is the (mass) fraction of molecular gas in the diffuse phase.
Observations of both low- and high-$z$ ULIRGs point to galaxy-wide
\aco{} values close to $0.8~\acounit{}$ \citep{ downes1998,papadopoulos2012b,
magdis2011, spilker2015}, consistent with the decrease
in \aco{} expected for a galaxy with a large fraction of its gas in a
diffuse phase.  On the other hand, the average \aco{} value for
molecular gas in local star-forming galaxies largely mirrors the Milky~Way
value \citep{sandstrom2013}, consistent with gas that primarily
resides in self-gravitating GMCs. 

By investigating the relationship between \aco{} and total mass
surface density for a large sample of galaxies at $z\sim0$ and
$z\sim1$, our study probes the evolution of molecular gas conditions
over cosmic time. In particular, we investigate the possibility that
the structure of molecular gas at $z\sim1$ is significantly different
from that observed at $z\sim0$, with more gas in large (kpc-scale)
star-forming clumps than self-gravitating GMCs, as suggested by
high-$z$ IFU observations \citep{forstershreiber2009, genzel2011,
  ceverino2010}.
In this case (high $f_d$), the stellar mass enclosed in these clouds
would raise their velocity dispersion, leading to a negative
correlation between \aco{} and \sigtot{} at $z\sim1$. However, if most
of the gas is in the form of GMCs similar to those in the Milky~Way
disk (low $f_d$), we expect a constant \aco{}, independent of
\sigtot{}.

\begin{figure*}
\centering
\includegraphics[width=1\linewidth]{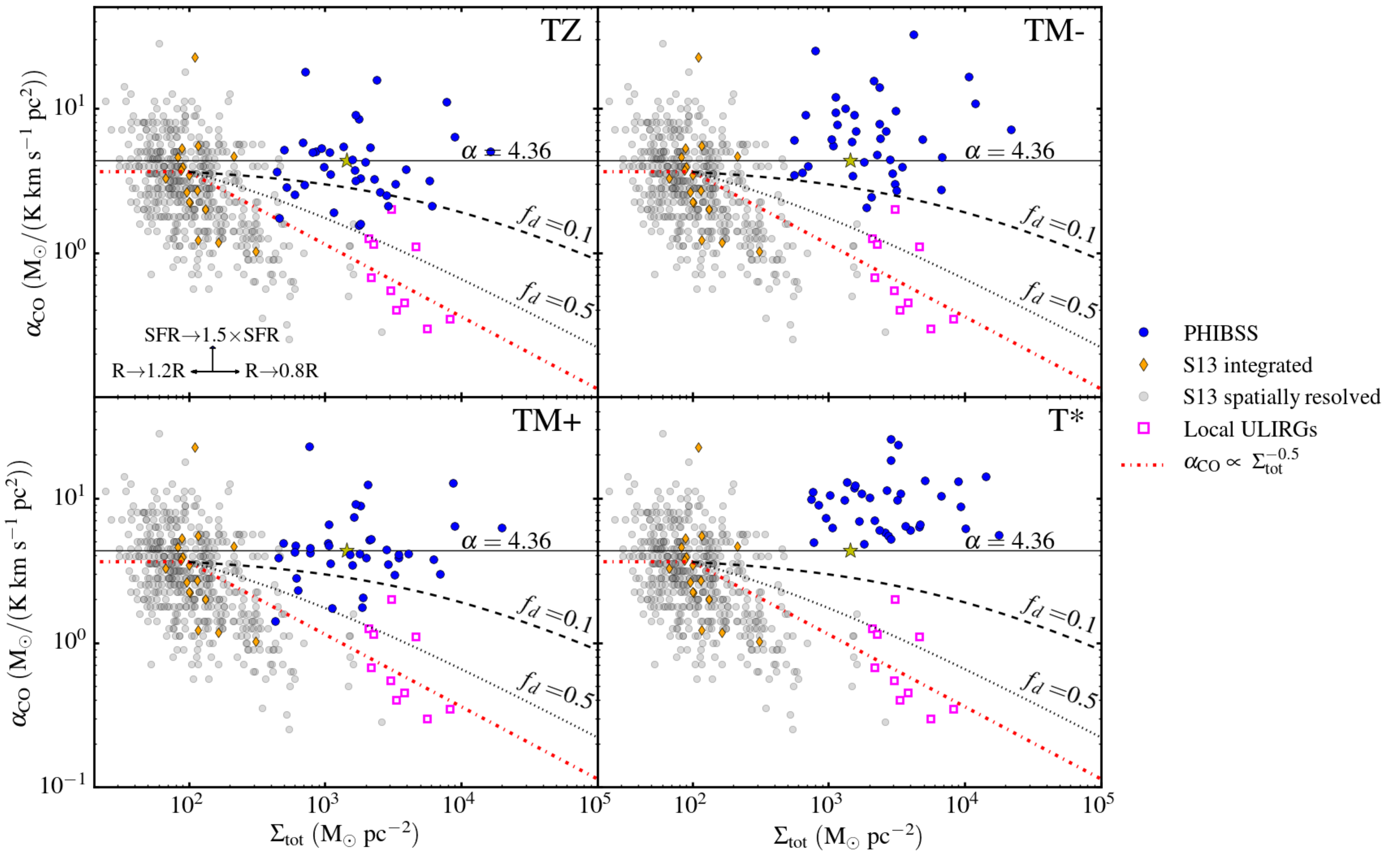}
\caption{The dependence of \aco{} on mass surface density for the
  PHIBSS sample. Blue circles represent our measurements of galaxies
  in PHIBSS, with the \tdep{} model used in each \aco{} calculation
  shown in the top of each panel, whereas grey, orange, and magenta
  points are the same as in
  Figure~\ref{fig:acodensity4plotcgas}. Vectors in the lower-left
  corner of the top-left panel indicate how a change in SFR or \re{}
  affect our measurements of \aco{} and \sigtot{}.  Motivated by the
  noticeable trend between \aco{} and surface density in the grey and
  magenta points, some models \protect \citep{narayanan2012,
    bolatto2013} suggest that \aco{} decreases smoothly with
  increasing surface density in extended molecular clouds. As in
  Figure~\ref{fig:acodensity4plotcgas}, we show the model from
  \protect \citet{bolatto2013}, Eq.~31, shifted to the metallicity of our high-$z$
  sample (following the redshift-dependent mass-metallicity relation of \citet{zahid2014})
   as the red dot-dashed line and include
  models for the dependence of \aco{} on surface density assuming a
  mix of diffuse molecular gas (with $\aco{}\propto\Sigma_{\rm
    tot}^{-0.5}$) and gas in Milky~Way-like GMCs (with a constant
  \aco{}; see Equation~ \ref{eqn:avgeqn}). The black dotted line
  represents a $50$-$50$ mix, by mass, of diffuse gas and GMC-like
  gas, and the black dashed line represents a mix of $10\%$ diffuse
  gas and $90\%$ GMC-like gas. The \aco{} points within the PHIBSS
  sample do not show a significant decrease with increasing surface
  density, suggesting that molecular gas is primarily in the form of
  self-gravitating GMCs as opposed to large star-forming clumps.}
\label{fig:acodensity4plotphibss}
\end{figure*}

\subsection{COLD GASS}

Figure~\ref{fig:acodensity4plotcgas} shows the relationship between
\aco{} and total mass surface density in our low-$z$
sample. Throughout, we define total mass surface density as the total
stellar \emph{and} gas mass for each system divided by the area within
$r_{e}$:
\begin{equation}
\Sigma_{\rm tot} = \frac{M_*+ M_{{\rm
      H}_2}+M_{\rm H\scalebox{0.55}{\rm I}}}{\pi r_{e}^{2}} \, . 
\end{equation}
When calculating $\Sigma_{\rm tot}$, $M_{{\rm H}_2}$ is inferred from the associated \tdep{} model, in
	order to be consistent with the rest of the analysis.
In our low-$z$ sample, stellar mass represents the dominant mass
component, with $M_{\rm gas}/(M_{\rm gas}+M_*)\sim22\%$ on average, 
such that we adopt the $r$-band Petrosian half-light radius
for $r_{e}$.
Our results remain unchanged, however, when taking $r_{e}$ to be the
$u$-band Petrosian half-light radius, which tracks the typically
more-extended light associated with star formation.
Alongside the COLD GASS sample in Fig.~\ref{fig:acodensity4plotcgas},
we also plot \aco{} values determined from fits to molecular emission
in local ULIRGs, assuming a single component ISM model, from \citet{papadopoulos2012}
and dust-based \aco{} measurements from \cite{sandstrom2013}, which track \aco{}
in kpc-sized regions in local ULIRGs and disk galaxies, respectively.
These measurements from the literature show that gas in the densest
regions of galaxies is affected by the galactic environment, such that
\aco{} decreases with increasing \sigtot{}. 
In contrast to these spatially-resolved measurements, however, the
system-wide (or integrated) \aco{} values for the \cite{sandstrom2013}
sample do not show a correlation between \aco{} and \sigtot{}, indicative of
a molecular gas component dominated by Milky~Way-like GMCs.
As a connection between our understanding of \aco{} in individual
molecular clouds and galaxy-wide \aco{} values,
Fig.~\ref{fig:acodensity4plotcgas} shows the expected trend between
\aco{} and \sigtot{} with $10\%$ and $50\%$ of gas (by mass, following
Equation~\ref{eqn:avgeqn}) residing in a diffuse phase (black dashed
and dotted lines, respectively) --- where we assume $\alpha_{G}=
\alpha_{\rm MW}$ shifted to the average metallicity of the low-$z$ sample
($2.6~\acounit{}$ independent of \sigtot{}), and
$\alpha_{D}\propto\Sigma_{\rm tot}^{-0.5}$, normalized to
$2.6~\acounit$ at $\sigtot{}=100~{\rm M_\odot}~{\rm pc^{-2}}$.

For the COLD GASS sample, which probes total mass surface densities
intermediate between that of local ULIRGs and the \cite{sandstrom2013}
sample, we
do not find a decrease in \aco{} with increasing
\sigtot{}. Independent of the depletion time model assumed, the
inferred \aco{} values are not significantly lower for the
high-density systems relative to their low-density counterparts.
An ordinary least-squares linear regression between \aco{} and
\sigtot{} reveals that none of the $t_{\rm dep}$ models yield a slope
more than $1\sigma$ below $0$.
The lack of a significant trend between \aco{} and total mass surface
density suggests that CO-emitting molecular gas within the COLD~GASS
sample is primarily in the form of self-gravitating GMCs similar to
those in the Milky~Way disk.

\subsection{PHIBSS}
\label{sec:phibssacodensity}

Figure~\ref{fig:acodensity4plotphibss} shows the relationship between
\aco{} and total mass surface density in our high-$z$ sample, where
\sigtot{} is again calculated as the total stellar and (molecular) gas
mass divided by the area within \re{}. Excluding the atomic gas
component in our high-$z$ sample probably does not significantly affect our
results, especially recognizing that H{\scriptsize I} represents
$\lesssim~10\%$ of the baryonic content in massive galaxies locally \citep{catinella2010}.
Although it traces bluer, slightly more extended emission than the
$r$-band measurements used in the low-$z$ sample, the $I$-band size is
employed in this calculation due to its high spatial resolution and
sensitivity.

As for the low-$z$ sample, we find no significant correlation between
\aco{} and \sigtot{}: a linear fit to the data shows that none of the
assumed \tdep{} models yield \aco{} values that decrease with
increasing surface density.
This result remains unchanged when utilizing complementary \ha{},
$J_{\rm F120W}$, or $H_{\rm F160W}$ size measurements to measure
\sigtot{} (see Sec.~\ref{sec:resolution}
for a discussion of how the distribution of molecular gas 
in our systems impacts our results).
Overall, the relationship between \aco{} and surface density in the
PHIBSS sample is discrepant with a scenario in which molecular gas in
high-$z$ galaxies is primarily in the form large, diffuse clouds, like
those found in high-density galaxy centers or merger-driven ULIRGs.
Model~\ref{tdepssfr} points are higher than those of our other models; however,
uncertainties in the redshift evolution of \tdep{} result in significant uncertainties in the normalization of
our high-$z$ \aco{} values (see Sec.~\ref{sec:tdepevolution}).
Values of \aco{} for our \ref{tdepssfr} model are greater than those of our other depletion time
models, in part due to differences in the star formation rate indicators used to
track the evolution of the sSFR (and thus the redshift-dependent normalization of the \ref{tdepssfr} model)
versus that used to measure the SFRs for our high-$z$ sample (note that
model~\ref{tdepssfr} has the highest average \tdep{} among our models at high-$z$; see Fig.~\ref{fig:tdepmodels}).
 An additional redshift dependence in model \ref{tdepssfr}, like that suggested in
 \citet[][$\tdep\propto(1+z)^{-0.3}$]{genzel2015}, would lower the model \ref{tdepssfr} \aco{} values, significantly reducing the discrepancy between the model~\ref{tdepssfr} points and the Milky~Way value.
Regardless, the model~\ref{tdepssfr} \aco{} values are consistent with
the non-correlation between \aco{} and \sigtot{} observed for the rest of the models.

As in Fig.~\ref{fig:acodensity4plotcgas}, black dashed and dotted lines
in Fig.~\ref{fig:acodensity4plotphibss} indicate the predicted correlation
between \aco{} and \sigtot{} for a sample with $10\%$ and $50\%$ of the
molecular gas in a diffuse phase respectively.
The PHIBSS data are consistent with $\lesssim10\%$ of the molecular
gas in typical massive star-forming galaxies residing in a diffuse
phase (contributing $\lesssim30\%$ of the CO luminosity),
with the vast majority of gas contained within self-gravitating GMCs.

The lack of an observed decrease in \aco{} for high-density
star-forming galaxies at $z\sim1$ is arguably more surprising than the
same result at $z\sim0.05$.
The high-$z$ sample is much more dense on average than the low-$z$
sample ($\langle\Sigma_{\rm tot}\rangle=1600~{\rm M_\odot~pc^{-2}}$
for PHIBSS compared with $\langle\Sigma_{\rm tot}\rangle=800~{\rm
  M_\odot~pc^{-2}}$ for COLD~GASS) and contains a higher fraction of
highly star-forming ULIRGs than the low-$z$ sample, both thought to be
indicative of low \aco{} values. Regardless of the depletion time
model adopted, however, most ($92-100$ per-cent) of the \aco{} measurements are
closer to the Milky Way value ($4.36~\acounit$) than the canonical
ULIRG value ($0.8~\acounit$).  Despite the changing ISM conditions
between $z\sim0$ and $z\sim1$, the trend between \aco{} and density in
the high-$z$ sample follows the invariance between \aco{} and surface
density observed in the COLD~GASS sample locally, which suggests that
the molecular gas in typical star-forming galaxies at $z\sim1$ is
primarily in the form of self-gravitating GMCs instead of large, diffuse clouds.

\section{Discussion}
\label{sec:discussion}

Across a broad range of assumed molecular depletion times, we do not
observe a significant correlation between \aco{} and total mass
surface density for our sample of massive star-forming galaxies at
$z\sim0$ and $z\sim1$.
Instead, our analysis indicates that objects in our sample primarily
contain molecular gas in the form of self-gravitating GMCs as opposed to in a
diffuse phase. 
This result is somewhat surprising, particularly at high-$z$, where observations
suggest that kpc-scale clumps contribute substantially ($\sim20\%$)
to the galaxy-wide SFR \citep{forsterschreiber2011,genzel2011,wuyts2012} and
potentially form stars in a similar process to that observed in local ULIRGs.
Rather, we find that the molecular gas in typical star-forming systems, at both low- \emph{and} high-$z$, is primarily
in the form of self-gravitating molecular clouds. In this section, we investigate possible systematic
effects that could impact our results.

\subsection{Temperature and Excitation Effects}

If the high-density galaxies in our sample have systematically lower
ISM temperatures, the molecular gas would be less luminous despite its
elevated dispersion, thereby increasing \aco{} so as to potentially
flatten an existing correlation between \aco{} and \sigtot{}.
Observations of cold gas in star-forming galaxies with
\textit{Herschel}, however, indicate that gas temperatures remain
remarkably constant between $z\sim1$ and $z\sim0$ in typical systems
\citep{dunne2011, elbaz2010,magnelli2014,bethermin2015}.
Moreover, observations of mergers and ULIRGs typically yield gas
temperates elevated by a factor of $2-4$ (not lower) relative to that
measured in normal star-forming systems \citep{krumholz2007,
  papadopoulos2012b, kamenetzky2014}.
Such elevated gas temperates would produce lower \aco{} values, so as
to strengthen a potential correlation between \aco{} and surface
density, not weaken it. 

Furthermore, given that the brightness temperature ratio
$L^\prime_{32}/L^\prime_{10}(r_{31})=0.5$ is directly linked to the
inferred CO~$J$=($1\rightarrow0$) luminosity, any systematic variation
of $r_{31}$ across our $z\sim1$ sample would directly influence our
\aco{} observations.
Although the excitation temperature is linked with gas density on the
scales of individual clouds, global $r_{31}$ values exhibit little
variation with integrated galaxy properties. The value of $r_{31}=0.5$
has been observed in both normal galaxies \citep{mauersberger1999,
  yao2003} and local ULIRGs \citep{papadopoulos2012, bauermeister2013,
  daddi2015}. While there is significant ($40\%$) scatter in measured
$r_{31}$ values, no strong correlation between $r_{31}$ and stellar
mass, star formation rate, gas mass, or dust temperature has been
observed. If CO is in a more thermalized state in the densest systems,
the $r_{31}$ ratio will decrease, resulting in lower inferred \aco{}
values based on the inferred molecular gas mass. Although there is some
evidence for a higher $r_{31}$ in denser galaxies \citep{daddi2015},
and observations of two $z\sim2$ PHIBSS galaxies measure $r_{31}$ of
close to unity \citep{bolatto2015}, the thermalized limit of
$r_{31}=1$ restricts this effect to a factor of $2$, not enough to
fully explain the factor of $\sim5$ discrepancy between the PHIBSS
\aco{} values and those in similarly dense local ULIRGs.

\subsection{Resolution Effects}
\label{sec:resolution}

Another explanation for the results in Section~\ref{sec:acodensity} is
that our measurements of \sigtot{} are not accurate representations of
the molecular gas environment within our galaxy sample. Previous
studies that find a reduction in \aco{} with increasing \sigtot{}
resolve kpc-scale regions within nearby galaxies \citep{sandstrom2013,
  papadopoulos2012b}, whereas our measurements of \aco{} and \sigtot{}
are integrated over the entire system. As a result, our \sigtot{}
determination would fail to describe molecular gas conditions in a
galaxy with significantly offset molecular gas and stellar mass
distributions.

Observations of nearby galaxies, however, find that the distributions
of molecular gas and stellar mass are remarkably similar. For example,
\cite{regan2001} and \cite{leroy2008} construct high-resolution maps
of CO emission in $15$ and $23$ nearby galaxies, respectively; both
find that the CO scale length is consistent with the stellar scale
length for a wide range of galaxy types.
\cite{tacconi2013} arrive at the same result (albeit with
lower-resolution observations) for $21$ high-$z$ star-forming galaxies
with spatially-resolved CO maps from PHIBSS, including a subset of the
sources in our high-$z$ sample \citep[see also][]{genzel2013,
  freundlich13}.
The molecular gas distributions within nuclear starbursts, such as
those observed in local ULIRGs, on the other hand, clearly do not
match their stellar distributions.
Further verification that molecular gas at intermediate redshift does
not live in preferentially dense environments is possible via
high-resolution observations with ALMA, which -- in its most extended
configuration -- is capable of resolving the
$\mathrm{CO}$~$J$=($3\rightarrow2$) emission on $\sim{\rm kpc}$ scales
at $z\sim1$.
A high-resolution census of the molecular gas in typical star-forming
systems with ALMA (or NOEMA) would provide a better measurement of the
total mass surface density within molecular clouds, thereby serving as
a valuable check of our results.

\begin{figure*}
\centering
\includegraphics[width=1\linewidth]{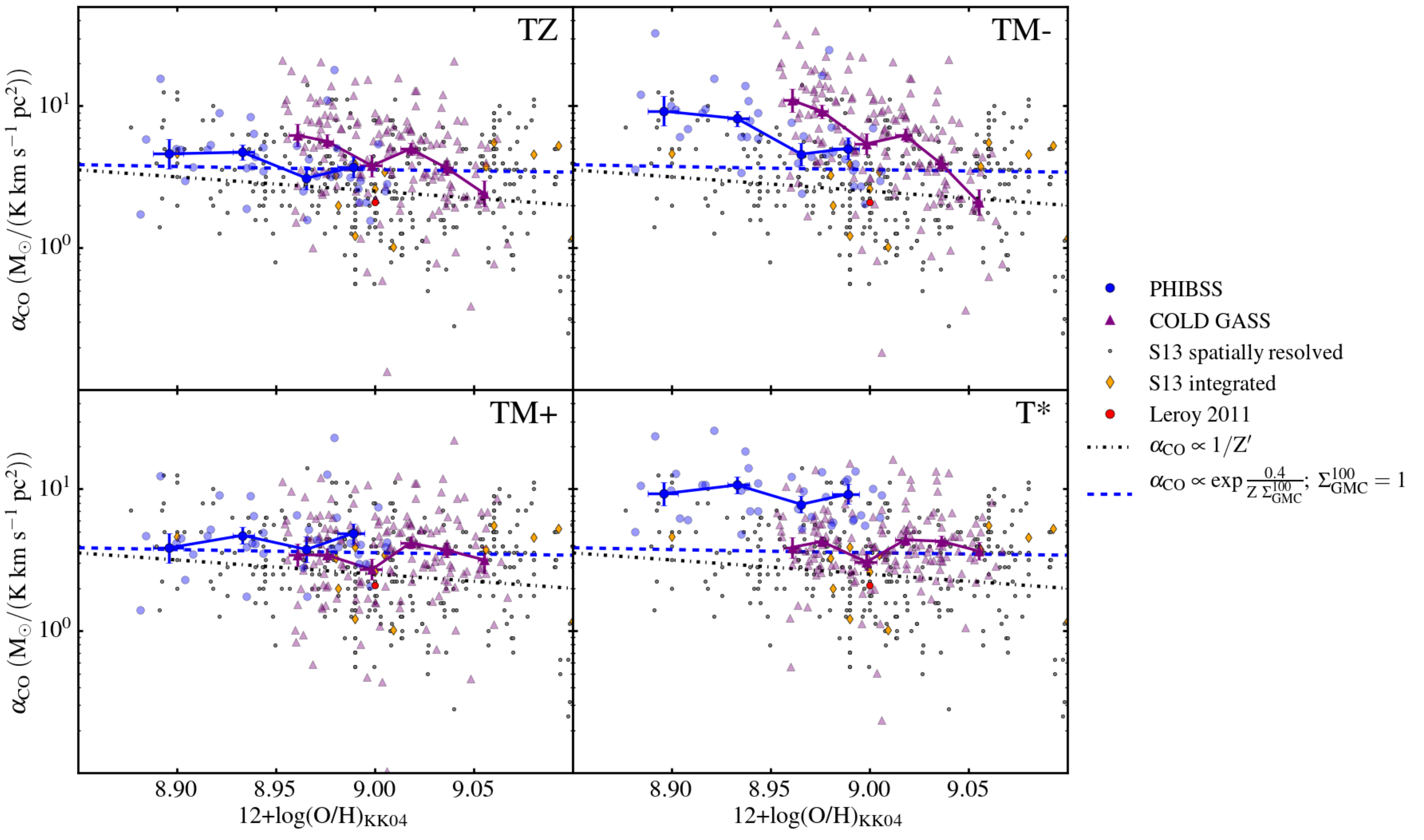}
\caption{The dependence of \aco{} on metallicity in both the low- and
  high-$z$ samples in our study. Blue, purple, grey, and orange points
  are the same as in Figures~\ref{fig:acodensity4plotcgas} and
  \ref{fig:acodensity4plotphibss}. Red points are taken from
  measurements within the Local Group from
  \protect\cite{leroy2011}. Two models for the dependence of \aco{} on
  metallicity are also shown: the blue dashed line shows the weak
  dependence of \aco{} on metallicity given in
  \protect\citet{bolatto2013}, and the black dot-dashed line shows the
  stronger dependence given by \protect\cite{narayanan2012}. Both the
  low- and high-$z$ samples qualitatively match the expected
  relationship between \aco{} and metallicity in all \tdep{} models
  except \ref{tdepdave}. Additionally, the scatter in our measurements
  at fixed metallicity is similar to that of
  \protect\cite{sandstrom2013}.}
\label{fig:acometal4plot_2z15}
\end{figure*}

\subsection{Metallicity Effects}
\label{sec:metallicity}
As discussed in Section~\ref{sec:introduction}, along with total
surface mass density, metallicity is the galaxy property most associated
with variation in \aco{}.
The dependence of \aco{} on metallicity is driven by the underlying
correlation between the fraction of CO-dark gas within a molecular
cloud and extinction, which in turn depends on metallicity such that
as gas-phase metallicity increases, \aco{} decreases.
Different parameterizations exist describing the dependence of \aco{}
on metallicity. 
For example, \cite{narayanan2012} arrive at $\aco{}\propto 1/Z^\prime$
as a fit to hydrodynamic simulations, while \citet{bolatto2013},
following the analytic arguments of \cite{wolfire2010}, give
$\aco{}\propto\exp(0.4/Z^\prime)$, where $Z^\prime$ is the
metallicity as a fraction of solar. Both of these parameterizations
provide a good fit to observations of \aco{} and gas-phase metallicity
which indicate that on average \aco{} decreases by roughly $1$~dex
over the range $8<12+\log({\rm O/H})<9$.

Since neither direct-$T$ nor strong-line metallicity measurements are
available throughout our sample, we estimate the gas-phase metallicity
of each galaxy in PHIBSS and COLD~GASS according to the observed
stellar mass, using an assumed
mass-metallicity relation. Metallicities are first determined from the
redshift-dependent mass-metallicity relation of \citet[][]{zahid2014},
and then converted to the metallicity scale of \cite{kobulnicky2004}
using the conversion from \cite{kewley2008}. This conversion allows us
to compare our results with measurements of local disks in the
\cite{sandstrom2013} sample and local dwarfs as measured by
\cite{leroy2011}.

Figure~\ref{fig:acometal4plot_2z15} shows the relationship between
\aco{} and metallicity in both the low- and high-$z$ samples.
Overall, our measurements are in line with the weak negative
correlation between \aco{} and metallicity predicted.
While we are unable to disentangle the contributions of $Z^\prime$ and
\sigtot{} in setting \aco{} for our observed galaxy sample, the
expected dependence of \aco{} on $Z^\prime$ suggests that metallicity
has a limited effect on our results, and in particular, would not
cancel an existing trend between \aco{} and \sigtot{}.
High-density galaxies tend to be more massive, and thus more metal
rich, with lower \aco{} values than their low-metallicity
counterparts.
Therefore, any underlying dependence of \aco{} on $Z^\prime$ would
serve to strengthen any existing anti-correlation between \aco{} and
\sigtot{}.
Moreover, the relatively weak dependence of \aco{} on metallicity in
the metal-rich regime, combined with the small range of metallicities
spanned by our sample, suggests that this effect should lower \aco{}
by at most $0.13$~dex across our high-$z$ sample and $0.12$~dex across
our low-$z$ sample.
This small decrease is certainly less important in setting \aco{} than
the decrease in \aco{} expected for a galaxy with a large fraction of
its gas in a ULIRG-like diffuse cloud.
While the presence of low-metallicity gas can significantly alter \aco{}, the
magnitude and direction of this effect in our sample are 
incompatible with a scenario in which the results in Sec.~\ref{sec:acodensity}
are explained as the combination of a density effect and a metallicity effect.

Given the range of stellar mass dependencies among our depletion time
models, it is unsurprising that we observe varying relationships
between \aco{} and metallicity in
Fig.~\ref{fig:acometal4plot_2z15}. For depletion time
models~\ref{tdepzmpos} and \ref{tdepssfr}, we see no significant
correlation between \aco{} and metallicity. For model~\ref{tdepz},
there is a slight negative trend between \aco{} and metallicity
($\log\aco{}-\log Z^\prime$ slope $=-3.4\pm 0.9$), while
model~\ref{tdepdave} results in a very steep relationship between
\aco{} and metallicity (slope $=-6.5\pm0.9$). 
	By inferring $Z^\prime$ from the observed stellar mass,
	our metallicity estimates are highly uncertain, so as to potentially
	smear out an underlying $\aco{}-Z^\prime$ trend. However, use of strong-line
	metallicity measurements, available for a subset of our low-$z$ sample
	\citep{tremonti2004}, results in a generally \emph{flatter} $\aco{}-Z^\prime$
	relation with comparable scatter.

As shown in Fig.~\ref{fig:acometal4plot_2z15}, all depletion time
models, except for \ref{tdepdave}, are consistent with the weak
negative correlation between \aco{} and metallicity observed locally
in the relevant metallicity range. While the points for model
\ref{tdepssfr} are higher at $z\sim1$ than $z\sim0$, their slope is
still consistent with the expected $Z^\prime$ dependence. The nature
of this general evolution in \aco{} is unclear. As discussed in
Section~\ref{sec:tdepevolution}, uncertainties in the redshift
evolution of \tdep{} result in uncertainties in the normalization of
our high-$z$ \aco{} values. 
Relative differences in \aco{} as a function of galaxy properties,
however, are not subject to this same uncertainty.
Regardless, the strong anti-correlation we find for
model~\ref{tdepdave} is inconsistent with the expected $Z^\prime$
dependence, such that model \ref{tdepdave} is incompatible with the
COLD GASS and PHIBSS observations. Alternatively, the weak correlation
between \aco{} and metallicity in models \ref{tdepz}, \ref{tdepzmpos},
and \ref{tdepssfr}, corresponding to a decrease in \aco{} of around
$0.2$~dex across our sample, is consistent with local observations and
models, providing further evidence in support of these depletion time
models.

\begin{figure}
\centering
\includegraphics[width=1\linewidth]{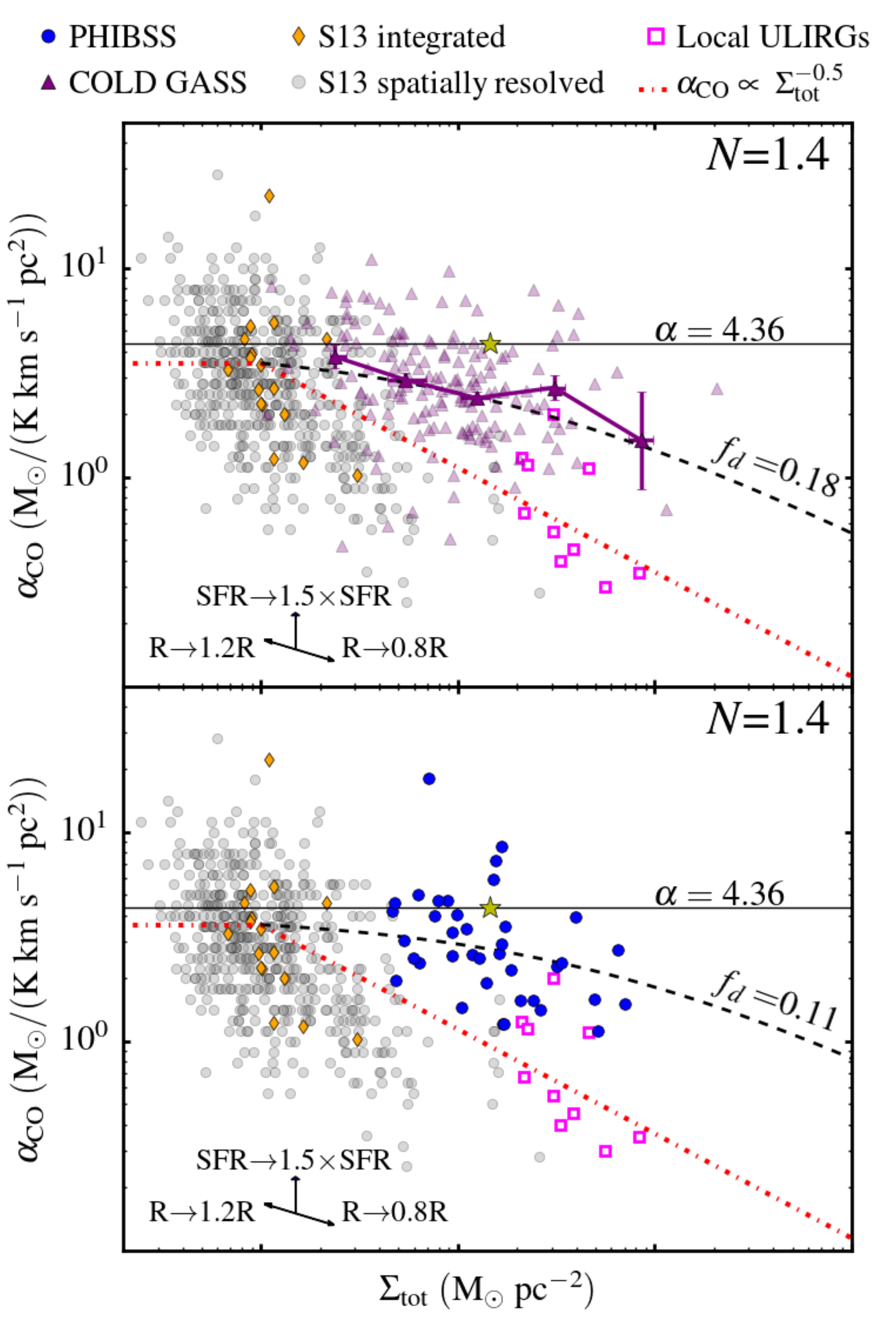}
\caption{The relationship between \aco{} and surface density assuming
  an $N=1.4$ Kennicutt-Schmidt law. Blue, purple, grey, and orange
  points are the same as in Figures~\ref{fig:acodensity4plotcgas} and
  \ref{fig:acodensity4plotphibss}. As in
  Figures~\ref{fig:acodensity4plotcgas} and
  \ref{fig:acodensity4plotphibss}, the model from \protect
  \citet{bolatto2013} is illustrated as the red dot-dashed line, and
  the expected trends between \aco{} and \sigtot{} for galaxies
  containing $18\%$ and $11\%$ diffuse gas are shown as the black
  dashed lines at low and high $z$ respectively. Adopting this surface-density
  dependent depletion time, a strong correlation between \aco{} and
  \sigtot{} is observed, consistent with a non-negligible fraction of
  molecular gas in the diffuse phase. The relationship between \aco{}
  and surface density does not change between the $z\sim1$ and
  $z\sim0$ samples, however. Thus, even if the depletion time is
  strongly correlated with surface density, we find no evidence of
  that the structure of molecular gas within `typical' star-forming
  galaxies is significantly different at $z\sim1$ compared with
  $z\sim0$.}
\label{fig:acodensityplots_ks}
\end{figure}

\subsection{Surface Density Dependent $\mathbf{\emph{t}_{dep}}$ Models}
\label{sec:tdepsigma}
While the depletion time models we adopt do not necessarily preclude a
correlation between \tdep{} and total mass surface density, some
studies report a more direct relationship between depletion time and
surface density. For example, a Kennicutt-Schmidt (KS) law
$(\Sigma_{\rm SFR}\propto\Sigma_{\rm mol~gas}^N)$ with $N=1.4$
\citep{kennicutt1998} or a model relating \tdep{} to the orbital time
of the gas \citep{kennicutt1998,kennicutt2007} or stellar surface
density \citep{boselli2014} all imply strong correlations between
\tdep{} and total mass surface density. 
Given that our \aco{} estimates are determined directly from
depletion time models (Eq.~\ref{eqn:acoeqn}), a model that assumes
a correlation between \tdep{} and surface density will undoubtedly
imply a trend between \aco{} and total mass surface density.

To demonstrate the extent of the effect produced by adopting one of
these models, we show the relationship between \sigtot{} and \aco{}
determined from an $N=1.4$ KS law in
Fig. \ref{fig:acodensityplots_ks}. In this case, a stronger
relationship between \aco{} and surface density is observed,
consistent with a $\sim20\%$ diffuse fraction at low-$z$ and a
$\sim10\%$ diffuse fraction at high-$z$ (contributing $\sim 41\%$ and $\sim 24\%$
to the CO luminosity at low- and high-$z$ respectively). The small change between these
results and the results presented in Sec.~\ref{sec:acodensity} is likely due to the fact that
molecular gas (controlling the density effect) represents a sub-dominant mass component within our systems.
 If $N$ is chosen to be less
than $1.4$ (in line with more recent results,
e.g.~\citealt{genzel2010}), the relationship between \aco{} and
\sigtot{} flattens, better matching the trend expected for a
$\lesssim10\%$ diffuse fraction for both samples.  While the adoption
of an $N>1.4$ KS law would increase the inferred diffuse fraction,
most measurements of $N$ within nearby galaxies lie between $1$
\citep{leroy2008,bigiel2008,bigiel2011,schruba2011} and $1.4$
\citep{kennicutt1998,krumholz2009,kennicutt2012}, and CO-independent
measurements within local galaxies find $N$ to be consistent with $1$
\citep{leroy2013}. Furthermore, if we take \tdep{} to be proportional
to the orbital time of the gas \citep{kennicutt1998}, determined from
the H{\sc i} line width, the relationship between \aco{} and \sigtot{}
matches the trend observed using the $N=1.4$ KS law in the low-$z$
sample. In the high-$z$ sample, H{\sc i} orbital velocities are not
available and there is no evidence that the CO line width is dominated
by clear rotational motions, so we do not test this case.  Although
the adoption of a \tdep{} model that strongly correlates with surface
density results in a stronger correlation between \aco{} and \sigtot{}
within our sample, the relationship between \aco{} and total mass
surface density, indicative of the overall structure of molecular gas
within galaxies, does not change between the low-$z$ and high-$z$
samples.  Specifically, our analysis indicates that molecular gas,
both at low- and high-$z$, is primarily distributed among
self-gravitating GMCs and not large, diffuse clouds.

\subsection{Evolution of  $\mathbf{\emph t}_{\mathbf {dep}}$}
\label{sec:tdepevolution}

The nature of the star-formation process at high-$z$, which is
directly connected to the $z\sim1$ depletion time, is the subject of
considerable debate. We have adopted forms of \tdep{} that evolve
slowly with redshift, consistent with qualitatively non- (or weakly-)
evolving star-formation process, motivated by many different CO-based
observations \citep{bauermeister2013b, geach2011, saintonge2013,
  magdis2012b, genzel2015}.
Specifically, \citet{genzel2015}, using both dust- and CO-based
gas measurements, find that depletion times evolve weakly with redshift
for galaxies on the star-forming main sequence.
In contrast, \cite{scoville2016}, using
stacked far-IR photometry of galaxies at $z > 1$, measure depletion
times on the order of $10^8$~yrs, at least a factor of $5$ below the
$z\sim1$ depletion times that we adopt. Such a remarkably short
depletion time implies that star formation at $z\sim1$ is more similar
to that occurring in local ULIRGs than in the Milky~Way.

Our primary results, which show \aco{} to be independent of \sigtot{} at
$z\sim0$ and $z\sim1$, are largely insensitive to the assumed
evolution of \tdep{}.
Whereas depletion times as low as $10^8$~yrs result in \aco{}
values close to those of local ULIRGs, they only affect the normalization
of our measurements, leaving the lack of a significant
trend between \aco{} and \sigtot{} unexplained.
On the other hand, the motivation for lowering \aco{} in
high-surface-density galaxies at $z\sim1$ may not apply for our sample. 
While kpc-scale
clumps of star-formation activity have been observed in high-$z$
galaxies \citep{forsterschreiber2011, genzel2011}, it is possible that
these star-forming clumps are composed of a collection of GMCs
\citep[see][]{hemmati2014}.
In addition, although a variety of star formation histories have been shown 
to be consistent with the observed star-forming main sequence 
\citep{kelson2014,abramson2015}, depletion times as low as $10^8$ years
require an extremely unlikely conspiracy of gas accretion and cooling
to avoid conflict with the low scatter in the main sequence at $z>1$ \citep{speagle2014}.
As we do not observe a trend between \aco{} and \sigtot{}, our
observations indicate that molecular gas in typical star-forming
galaxies at $z<1.5$ largely resides in self-gravitating GMCs with a 
Milky~Way \aco{} value. This molecular gas structure, together with the high-$z$
CO observations, implies a weakly evolving \tdep{} across cosmic time.

\section{Conclusions}
\label{sec:conclusions}

For a sample of $164$ low-$z$ and $38$ high-$z$ star-forming galaxies
with CO detections drawn from COLD~GASS and PHIBSS, respectively, we
estimate the CO-H$_{2}$ conversion factor (\aco{}) using the observed
star formation rate and an adopted depletion time (\tdep{}) to infer
the molecular gas mass.
In particular, we study the relationship between \aco{} and total mass
surface density (\sigtot{}), constraining the structure of molecular
gas over cosmic time.
Our primary conclusions are as follows:

\begin{itemize}[leftmargin=0.25cm]

\item For a broad range of assumed depletion times, we do not find a
  significant correlation between \aco{} and total mass surface
  density for both our low-$z$ and high-$z$ samples, which suggests
  that $\lesssim10\%$ of the molecular gas (by mass; $\lesssim30\%$ by luminosity) in these systems
  is contained in diffuse clouds, akin to those that populate local
  ULIRGs. 
  Instead, we find that the molecular content of typical star-forming
  galaxies at $z < 1.5$ is primarily comprised of self-gravitating
  GMCs, with an \aco{} value comparable to that found in the Milky
  Way.\\

\item For both our low-$z$ and high-$z$ samples, which primarily
  include metal-rich systems, we find that the relationship between
  \aco{} and gas-phase metallicity is consistent with the weak
  negative correlation observed locally. As this relationship is
  driven by gas chemistry within molecular clouds, the constancy of
  this relationship is evidence that the small-scale physics of GMC
  collapse at $z\sim1$ is similar to that locally, despite potential
  differences in large-scale ISM conditions. \\ 

\item Altogether, our analysis points to a molecular gas depletion
  time that is weakly evolving with redshift, such that typical
  star-forming galaxies at $z>1$ are more similar to scaled-up Milky
  Way-like systems than local ULIRGs. \\

\end{itemize}

\section*{Acknowledgements}

We are grateful to the anonymous referee, whose comments significantly improved the 
clarity of this work.

Support for this work was provided by NASA through grants (GO-12547
and AR-13242) from the Space Telescope Science Institute, which is
operated by the Association of Universities for Research in Astronomy,
Inc., under NASA contract NAS 5-26555.
This work was also supported, in part, by NSF grant AST-1518257.
TC and MCC thank the National Achievement Rewards for College
Scientists (ARCS) Foundation and the International Space Science
Institute (ISSI), respectively, for support of this work.
The observations presented here would not have been possible without
the diligence and sensitive new generation receivers from the IRAM
staff - for this they have our highest admiration and thanks. We also
thank the astronomers on duty and telescope operators for delivering
consistently high quality data to our team.

Funding for SDSS-III has been provided by the Alfred P. Sloan Foundation, the Participating Institutions, the National 
Science Foundation, and the U.S. Department of Energy Office of Science. The SDSS-III web site is 
http://www.sdss3.org/.

SDSS-III is managed by the Astrophysical Research Consortium for the Participating Institutions of the SDSS-III 
Collaboration including the University of Arizona, the Brazilian Participation Group, Brookhaven National Laboratory, 
Carnegie Mellon University, University of Florida, the French Participation Group, the German Participation Group, 
Harvard University, the Instituto de Astrofisica de Canarias, the Michigan State/Notre Dame/JINA Participation Group, 
Johns Hopkins University, Lawrence Berkeley National Laboratory, Max Planck Institute for Astrophysics, Max 
Planck Institute for Extraterrestrial Physics, New Mexico State University, New York University, Ohio State University, 
Pennsylvania State University, University of Portsmouth, Princeton University, the Spanish Participation Group, 
University of Tokyo, University of Utah, Vanderbilt University, University of Virginia, University of Washington, and 
Yale University.

This research has made use of the NASA/IPAC Extragalactic Database (NED) which is operated by the Jet Propulsion 
Laboratory, California Institute of Technology, under contract with the National Aeronautics and Space 
Administration.

This research made use of {\texttt{Astropy}}, a community-developed
core Python package for Astronomy \citep{astropy}. Additionally, the
Python packages {\texttt{NumPy}} \citep{numpy}, {\texttt{iPython}}
\citep{ipython}, {\texttt{SciPy}} \citep{scipy}, and
{\texttt{matplotlib}} \citep{matplotlib} were utilized for the
majority of our data analysis and presentation.
This work has made use of the Rainbow Cosmological Surveys Database,
which is operated by the Universidad Complutense de Madrid (UCM),
partnered with the University of California Observatories at Santa
Cruz (UCO/Lick, UCSC), as well as the NASA/IPAC Extragalactic Database
(NED) which is operated by the Jet Propulsion Laboratory, California
Institute of Technology, under contract with the National Aeronautics
and Space Administration.

\bibliography{acopaper}

\end{document}